\RequirePackage{booktabs}
\documentclass[sn-mathphys,Numbered]{sn-jnl}

\usepackage{graphicx}%
\usepackage{multirow}%
\usepackage{amsmath,amssymb,amsfonts,cancel}%
\usepackage{amsthm}%
\usepackage{mathrsfs}%
\usepackage[title]{appendix}%
\usepackage{xcolor}%
\usepackage{textcomp}%
\usepackage{manyfoot}%
\usepackage{booktabs}%
\usepackage{algorithm}%
\usepackage{algorithmicx}%
\usepackage{algpseudocode}%
\usepackage{listings}%
\usepackage{gensymb}
\usepackage{braket}
\usepackage{lscape}
\usepackage{multirow,multicol}
\usepackage{comment}
\usepackage{afterpage}
\usepackage{tikz}
 
\usepackage{geometry}

\theoremstyle{thmstyleone}%
\theoremstyle{thmstyletwo}%

\theoremstyle{thmstylethree}%

\allowdisplaybreaks

\def\wtil#1{\widetilde{#1}}

\def\tb{\bar{t}}
\def\ttz{t\bar{t}Z}
\def\l{\left}
\def\r{\right}

\begin{document}
	
	\title[Article Title]{On bipartite and tripartite entanglement at present and future particle colliders}
	
	
	\author*[1]{\fnm{Amir} \sur{Subba}}\email{as19rs008@iiserkol.ac.in}
	
	\author[2]{\fnm{Rafiqul} \sur{Rahaman}}\email{rafiqul@if.usp.br}
	\equalcont{These authors contributed equally to this work.}
	
	
	\affil*[1]{\orgdiv{Department of Physical Sciences}, \orgname{Indian Institute of Science Education and Research Kolkata}, \orgaddress{\street{Campus Road}, \city{Mohanpur}, \postcode{741246}, \state{West Bengal}, \country{India}}}
	
	\affil[2]{\orgdiv{Instituto de F\'isica}, \orgname{Universidade de S\~ao Paulo}, \orgaddress{\street{Rua do Mat\~ao 1371, Butanta}, \city{S\~ao Paulo}, \postcode{10587}, \state{SP}, \country{Brazil}}}

	
	\newcommand{\RR}[1]{\textcolor{magenta}{\bf(RR: #1)}}
	
	\abstract{
Entanglement, rooted in the non-deterministic, non-local nature of quantum mechanics, serves as a fundamental correlation. High-energy particle colliders offer a unique platform for exploring entanglement in the relativistic regime. The recent observation of entanglement in $t\bar{t}$ production by ATLAS has sparked significant interest in investigating entanglement phenomena at colliders. While bipartite entanglement receives extensive attention, tripartite entanglement remains relatively uncharted. We investigate tripartite entanglement in $t\bar{t}Z$ production at the Large Hadron Collider (LHC) within the Standard Model and with a dimension-$8$ effective operator. Additionally, we explore bipartite entanglement in $t\bar{t}$, $tW^-$, and di-boson production processes, namely $W^+W^-$, $ZZ$, and $W^+Z$, at the LHC and future $e^+e^-$ collider. We numerically compute various measures of entanglement through Monte Carlo events based on the spin density matrix, with its elements (polarization and spin correlation) obtained by analyzing the angular distribution of the final decayed leptons. 
	}
	
	
	
	
	\maketitle
	
	\section{Introduction}\label{sec1}
 Einstein, Rosen, and Podolsky (EPR)  highlighted a phenomenon in quantum mechanics where two spatially separated events could be determined by observing only one of them and termed it as \emph{spooky action at a distance}~\cite{Einstein:1935rr} challenging classical intuitions about causality and locality. In their attempt to reconcile these seemingly paradoxical effects (EPR paradox), the authors introduced the notion of hidden variables to restore a more deterministic and local interpretation of quantum mechanics. However, Bell countered that predictions of quantum mechanics are incompatible with hidden variable theories through some inequality that may be violated in quantum mechanics but preserved in classical physics. The concept that events might be correlated beyond the simple classical sense formulated by Schrodinger in his famous cat in a box Gedanken experiment~\cite{Schrdinger1935DieGS} known as \textit{Quantum Entanglement} (QE).
 
 Bell's inequality was later reformulated into a more experimentally testable form known as the Clauser-Horne-Shimony-Holt (CHSH) inequality~\cite{Clauser:1969ny}. The CHSH inequality is given as~\cite{Clauser:1969ny},
 \begin{equation}
 	|\langle ba \rangle + \langle ba^\prime \rangle + \langle b^\prime a \rangle - \langle b^\prime a^\prime \rangle| \le 2,
 \end{equation}
 where $a,$ and $a^\prime$ are the measurement setting on apparatus Alice ($A$), while $b,$ and $b^\prime$ are adjustable parameters in apparatus Bob ($B$). 
 The CHSH inequality can be violated by quantum mechanics as follows: consider $\vec{a} = \vec{\sigma}_A\cdot \hat{a},\vec{a}^\prime = \vec{\sigma}_A\cdot \hat{a}^\prime$, $\vec{b}=\vec{\sigma}_B\cdot \hat{b}$, and $\vec{b}^\prime = \vec{\sigma}_B\cdot \hat{b}^\prime$ acting on Alice and Bob qubit. If Alice and Bob share maximally entangled state $\ket{\psi}$ then, $\bra{\psi}(\vec{\sigma}_A\cdot \hat{a})(\vec{\sigma}_b\cdot \hat{b})\ket{\psi} = -\cos\theta$, where $\theta$ is the angle between two unit vectors $\hat{a}$, and $\hat{b}$. Further, we may assume the four-unit vectors to be co-planar and separated by  $45^\degree$ angles, providing
 \begin{equation}
 	\langle \vec{a}\vec{b} \rangle = \langle \vec{b}\vec{a}^\prime \rangle = \langle \vec{b}\vec{a}^\prime\rangle = -1/\sqrt{2}, \quad \langle \vec{a}^\prime \vec{b}^\prime \rangle = -\cos(3\pi/4)=1/\sqrt{2},
 \end{equation}
 which gives the CHSH inequality to be $2\sqrt{2} \le 2$, proving quantum mechanics can violate CHSH inequality. Several experiments~\cite{Freedman:1972zza,Aspect:1981zz,Aspect:1982fx,Weihs:1998gy,Clauser:1974tg,Tittel:1998ja} successfully quantified the violation of Bell's inequality using the CHSH inequality, primarily by exploiting the polarization of photons. Moreover, the violation of the CHSH inequality has been demonstrated even when two qubits, represented by two spatially separated photons~\cite{Tittel:1998ja} or electrons~\cite{Hensen:2015ccp}, are separated by several kilometers. These violations have also been observed in various other systems, including superconductivity circuits~\cite{Storz:2023jjx} and atoms~\cite{Rosenfeld:2017rka}. Entanglement also lies at the heart of many key discoveries, ranging from quantum teleportation~\cite{Bennett:1992tv} to quantum dense coding~\cite{bennett1992communication}, quantum computation~\cite{PhysRevA.52.R2493,PhysRevLett.77.793,feynman2018simulating}, and quantum cryptography~\cite{PhysRevLett.67.661,Jennewein:2000zz}. Entanglement extends beyond distinguishable particles and manifests between various degrees of freedom as    well~\cite{PhysRevLett.95.260501,RevModPhys.84.777,PhysRevA.79.030301,PhysRevA.94.030304}.
 
 In particle physics, Bell's inequality violation was first investigated in low-energy proton-proton scattering~\cite{Lamehi-Rachti:1976wey}, and its feasibility was explored at $e^+e^-$ colliders in hadronic~\cite{Tornqvist:1980af} and $\tau$ leptons~\cite{Privitera:1991nz} final states. A discussion regarding observing Bell's inequality violation in the Large Electron–Positron Collider (LEP) was presented in Ref.~\cite{Abel:1992kz}. Several tests of entanglement have been suggested in high-energy experiments involving Positronium \cite{Acin:2000cs}, Charmonium \cite{Baranov:2008zzb} decays, and neutrino oscillations \cite{Banerjee:2015mha,Balantekin:2023qvm}. Bell's inequality violation has also been observed in neutral Kaon oscillations due to CP-violation effects \cite{Benatti:1997fr,Benatti:1999du,Bertlmann:2001ea,Banerjee:2014vga}.
 
 The majority of experiments aimed at testing the violation of local hidden-variable (LHV) predictions have been conducted at low-energy scales. Conversely, high-energy particle accelerators such as the Large Hadron Collider (LHC) produce particles in extreme relativistic regimes, thus naturally providing a platform for entanglement exploration. Owing to the significant opportunities provided by high-energy experiments, recently numerous studies dedicated to investigating entanglement among the final state particles can be found in various references \cite{Acin:2000cs,Baranov:2008zzb,Banerjee:2015mha,Balantekin:2023qvm,Benatti:1997fr,Benatti:1999du,Bertlmann:2001ea,Banerjee:2014vga,Go:2003tx,Bertlmann:2004cr,Caban:2008qa,Afik:2020onf,Fabbrichesi:2021npl,Barr:2021zcp,Gong:2021bcp,Severi:2021cnj,Larkoski:2022lmv,Afik:2022kwm, Barr:2022wyq,Aguilar-Saavedra:2022uye,Aoude:2022imd, Fabbrichesi:2022ovb,Ashby-Pickering:2022umy,Severi:2022qjy,Altakach:2022ywa,Fabbrichesi:2023jep,Bernal:2023ruk, Morales:2023gow,Aguilar-Saavedra:2022wam,Aguilar-Saavedra:2022mpg,Aguilar-Saavedra:2024vpd,Maltoni:2024tul,Aguilar-Saavedra:2024hwd,Aguilar-Saavedra:2024fig,Bernal:2023jba,Bi:2023uop,Li:2024luk,Ehataht:2023zzt,Cheng:2023qmz,Bernal:2023jba,Ma:2023yvd,Aoude:2023hxv,Dong:2023xiw,Afik:2022dgh,Wu:2024mtj}.
 The top quark being exceptionally massive, offers a unique avenue for testing various measures of entanglement at high energy collider. As it decays before
 hadronization its polarizations are preserved in its decay products and can be measured from the angular distributions of its decay products. The polarization and spin correlation (in case of $t\bar{t}$ production) are used to measure various measures of quantum entanglement. Several studies~\cite{Aoude:2022imd,Severi:2021cnj,Afik:2022kwm,Fabbrichesi:2022ovb,Fabbrichesi:2021npl,Afik:2020onf,Han:2023fci,Aguilar-Saavedra:2023hss,Varma:2023gwh,Mantani:2022dao,Severi:2022qjy,Dong:2023xiw,Ghosh:2023rpj,Altakach:2022ywa} have been conducted to probe the entanglement in top quark pair production. 
 Quantum entanglement has also been used to study new physics beyond the standard model (SM) in top quark pair production at the LHC~\cite{Fabbrichesi:2022ovb,Severi:2022qjy,Mantani:2022dao,Maltoni:2024tul}. On the experimental side, ATLAS collaboration observed for the first time the existence of entanglement in top quark pair production~\cite{atlascollaboration2023observation} at the LHC with a significance of more than $5\sigma$ near threshold energy with the formalism discussed in Ref.~\cite{Afik:2020onf}. 
 
 Besides entanglement and Bell's inequality, authors in Ref.~\cite{Afik:2022dgh} also talk about quantum discord and steering in top quark pair production. In addition to top quark pair production, 
 bipartite qubit entanglement has also been investigated in the decay of muon pairs~\cite{Aguilar-Saavedra:2023lwb}, as well as in the case of $2\otimes 3$ spin states in $W^\pm\gamma$~\cite{Morales:2023gow} and $tW^\pm$ production~\cite{Aguilar-Saavedra:2023hss}. In addition,  entanglement and Bell's inequality violation are also investigated in $\tau^+\tau^-$ production~\cite{Ehataht:2023zzt,Ma:2023yvd,Fabbrichesi:2022ovb,Altakach:2022ywa} and hyperon pair production~\cite{Gong:2021bcp,Wu:2024mtj}. In addition to the bipartite qubit (top quark pair, $\tau$ lepton pair), considerable attention has been given to studying quantum entanglement in bipartite weak bosons (qutrit) systems at particle colliders~\cite{Fabbrichesi:2023cev,Barr:2021zcp,Aguilar-Saavedra:2022wam,Bi:2023uop,Aoude:2023hxv,Barr:2021zcp,Aguilar-Saavedra:2022wam,Aguilar-Saavedra:2022mpg}. Fabbrichesi et al.~\cite{Fabbrichesi:2023cev} studied quantum entanglement within the electroweak sector, particularly focusing on scenarios such as Higgs boson decays to weak bosons ($H\to VV^\star,~V\in{W,Z}$) and diboson production ($ff^\prime \to VV^\prime$) at the LHC and future lepton colliders such as the International Linear Collider~(ILC)~\cite{ILC:2007oiw,ILC:2013jhg,Adolphsen:2013kya}. Their study revealed the violation of Bell's inequality in Higgs boson decays and diboson production at the LHC based on current experimental data. Moreover, the analysis of diboson entanglement was also conducted in the presence of new physics beyond the Standard Model (SM)~\cite{Bernal:2023ruk,Fabbrichesi:2023jep,Aoude:2023hxv}.
 However, experimental verification of the violation of Bell-type inequalities by weak bosons is yet to be established. While finalizing this article, a nice review came up by Barr and colleagues~\cite{Barr:2024djo}   
 on entanglement concerning $2\otimes 2$ and $3\otimes 3$ spin quantum states at colliders.
 
 After exploring bipartite entanglement, the next logical inquiry involves measuring entanglement in multipartite systems. Assessing entanglement in a system with more than two particles is typically more intricate compared to the bipartite scenario. In the case of a multi-qubit system, Mermin-Ardehali-Belinskii-Klyshko~\cite{PhysRevLett.65.1838,PhysRevA.46.5375,belinskiui1993interference} inequality can be employed to test the violation of LHV models' prediction by quantum mechanics. Quantum states that are not even bi-separable for all possible bi-partitions are called genuine multipartite entangled states. These entangled states are of special interest since they are the extreme version of entanglement, that is, all sub-systems contribute to the shared entanglement feature~\cite{RevModPhys.81.865,PhysRevA.79.062308,Guhne:2008qic}. Studies on multipartite entanglement can be found in Refs.~\cite{bengtsson2016brief,Enriquez_2016,Amico:2007ag,Walter:2016lgl,Cunha:2019jex,PhysRevApplied.18.034004,Solomon_2012,yanliu,article} and references therein. In this article, we will limit ourselves to the simplest multipartite system, i.e., the tripartite system. From the viewpoint of a maximally entangled state, the Greenberger-Horne-Zeilinger (GHZ) states~\cite{Greenberger1990-GREBTW-3} are the direct generalization of Bell's state from a bipartite to a tripartite system. The simplest example of a GHZ state involves three qubits and is represented by
 \begin{equation}
 	\ket{GHZ} = \frac{1}{\sqrt{2}}\left(\ket{000}+\ket{111}\right),
 \end{equation}
 and its other equivalent forms can be obtained under the local unitary transformations. The GHZ state correlations were first reported in the experiment with three photons~\cite{Bouwmeester:1998iz}. The Schmidt decomposition~\cite{schmidt1907theorie,eckert,PERES199516} for two-qubit system allows for one free parameter $\theta$ in 
 \begin{equation}
 	\ket{\psi} = \cos\theta\ket{00}+\sin\theta\ket{11},
 \end{equation}
 while for three-qubit system~\cite{PhysRevLett.85.1560} there exist five independent parameters in a generalized Schmidt decomposition. Thus, one single measure may not be sufficient to fully
 characterize the properties of multipartite entanglement~\cite{vidal}. Any tripartite state can be separated into four different classes: product state (complete separable), bi-separable states, GHZ, and W class ($
 \ket{W_3} = \frac{1}{\sqrt{3}}(\ket{100} + \ket{010} + \ket{001})
 $). In the former two classes, at least one qubit is disentangled from the rest of the system, and the latter two are genuine entangled states. Ma et.al~\cite{PhysRevA.83.062325} identify two conditions that have to be satisfied by any entanglement measure to be called a genuine multipartite entagnlement~(GME). The conditions are 
 \begin{itemize}
 	\item measure must be zero for all product and bi-separable states, and 
 	\item measure must be positive for all non-bi-separable states~(GHZ and W class).
 \end{itemize}
 There are many GME measures like the linear and nonlinear entanglement witnesses~\cite{PhysRevLett.104.210501,PhysRevLett.113.100501,PhysRevA.84.062306,PhysRevA.86.022319,PhysRevA.88.042328,PhysRevLett.111.110503,PhysRevLett.108.020502,PhysRevA.91.042339,PhysRevA.87.034301}, generalized concurrence~\cite{PhysRevA.83.062325,PhysRevA.85.062320,PhysRevA.86.062323,PhysRevLett.112.180501}, and Bell like inequalities~\cite{PhysRevLett.106.250404}. Li et.al.~\cite{article} obtained GME measure for an arbitrary tripartite quantum system based on positive partial transposition~(PPT)~\cite{Peres:1996dw} and computable cross norm criterion~(CCNR) or matrix realignment~\cite{OliverRudolph2003,chen2003matrix}. Xie and Eberly~\cite{PhysRevLett.127.040403} introduce the area of a concurrence triangle as a new measure for GME.
 
 A fundamental distinction between quantum entanglement and classical correlation lies in the limited ability to distribute quantum entanglement among multiple parties. In a scenario involving multiple parties, when two entities achieve maximum entanglement, they are unable to share this entanglement with any other component of the overall system which is referred to as the monogamy of entanglement~(MoE)~\cite{Terhal_2004,Yang_2006}. A MoE can be defined as $E(\rho_{A|BC}) \ge E(\rho_{A|B}) + E(\rho_{A|C})$ where $E(\rho_{A|BC})$ is an entanglement measure quantifying the degree of entanglement between sub-systems $A$ and $BC$, and $E(\rho_{A|B})(E(\rho_{A|C}))$ is the bipartite entanglement between $A$ and $B$~($A$ and $C$). This MoE for a three-qubit system was proved by Coffman, Kundu, and Wootters~(CKW)~\cite{PhysRevA.61.052306} using squared concurrence, and one can construct a GME for tripartite qubit with CKW inequality. CKW inequality holds only for $2\otimes 2\otimes 2$ system, which is violated for higher dimensional systems like $3\otimes 2\otimes 2$~\cite{Kim_2010} and $3\otimes 3\otimes 3$ quantum system~\cite{PhysRevA.75.034305}. For systems larger than $2\otimes 2\otimes 2$, MoE is obtained by convex-roof extended negativity~(CREN)~\cite{PhysRevA.79.012329}. In the high energy context, multipartite entanglement is under development.  In a recent study~\cite{sakurai2023threebody}, authors explored tripartite entanglement within the framework of high-energy particle colliders in particle decay generated by (pseudo)scalars, pseudo(vectors), and (pseudo)tensors exchanges.  A separate study~\cite{Aguilar-Saavedra:2024whi} explores
 tripartite entanglement among the two spins and total angular momentum in $H \to ZZ, WW$ decay. Another study~\cite{Morales:2024jhj} explores tripartite entanglement and Bell non-locality in the loop induced $H\to \gamma \ell^+\ell^-$ decays.

 In this article, we quantify the measure of entanglement for various high energy processes corresponding to $2\otimes 2$, $2\otimes 3, 3\otimes 3$, and $2\otimes 2\otimes 3$ state at the LHC and upcoming $e^-e^+$ collider. The analyses are performed based on the density matrix approach where the elements of the density matrix are measured from the decay angular distributions of the particles taking part in entanglement. 
 The structure of this article is as follows: in section~\ref{sec2}, we discuss various techniques and measures available to quantify entanglement in $2\otimes 2$, $2\otimes 3$, $3\otimes 3$, and $2\otimes 2\otimes 3$ quantum system. Section~\ref{sec:halfhalf} discusses the measure of entanglement in $t\bar{t}$ production process as an example of  $2\otimes 2$ quantum state at the LHC and future $e^-e^+$ Collider. Similarly, section~\ref{sec:halfone} deals for $2\otimes 3$ state entanglement, and it is probed in single top quark production associated with $W$ boson process at the LHC. Section~\ref{sec:oneone} describes the measure of entanglement for $3\otimes 3$ quantum state in $VV$ ($V=W^\pm,Z$)  production process at the LHC and  $e^-e^+$ Collider. In section~\ref{sec:halfhalfone}, we discuss the measure of tripartite entanglement with $2\otimes 2\otimes 3$ quantum state in $t\bar{t}Z$ production process at the LHC in the SM as well as in the presence of a dimension-8 effective operator. We conclude with further discussions in section~\ref{sec:conclude}.
\section{Measures of Entanglement}
\label{sec2}
In this section, we discuss different measures of entanglement associated with a quantum system. A quantum system $S$ is described in term of states $\ket{\psi}$ which are elements of a $n$-dimensional Hilbert space ${\cal H}_n$ (finite for simplicity) as
\begin{equation}
	\ket{S}=\sum_{i=1}^{n}\sqrt{p_i}\ket{\psi_i},~~p_i\geq 0,~~\sum_i^np_i=1.
\end{equation}
Expectation value of an operator $\hat{\cal O}$ in the quantum state will be given by
\begin{equation}
	\langle \hat{\cal O}\rangle=\sum_i p_i \bra{\psi_i}\hat{\cal O}\ket{\psi_i}.
\end{equation}
Statistically the quantum system $S$ can be described by a density matrix represented as
\begin{equation}
	\rho = \sum_i p_i \ket{\psi_i}\bra{\psi_i}.
\end{equation}
Expectation of $\hat{\cal O}$ will then be given by $\langle \hat{\cal O}\rangle={\rm Tr}[\rho\hat{\cal O}].$
The density matrix of the system holds the properties of hermiticity, i.e., $\rho^\dagger = \rho$, normalization (${\rm Tr}[\rho]=1$) and semi-definite positivity, i.e., $\langle\psi| \rho |\psi\rangle\geq 0$.  Additionally, ${\rm Tr}[\rho^2]\leq 1$, and its   eigenvalues, $\lambda_i$ satisfies $0<\lambda_i<1$ in general. The properties of the density matrix of a quantum system indicate whether the constituents of the systems are correlated quantum mechanically, i.e., whether quantum entanglement is present in the system. 
Below, we discuss various tests for the presence of entanglement as well as measures of entanglement in bipartite states with two qubits, a qubit and a qutrit, and two qutrits, and tripartite states with two qubits and a qutrit.

\subsection{Bipartite entanglement}
For a bipartite case defined within the Hilbert space ${\cal H}={\cal H}_A\otimes {\cal H}_B$, a state $\rho$ is considered separable if it can be expressed as a tensor product of individual sub-states~\cite{Werner:1989zz}:
\begin{equation}
	\rho = \sum_{ij} w_{ij} \rho_A^i \otimes \rho_B^j, \quad \sum_{ij} w_{ij} = 1, \quad w_{ij} \ge 0,
\end{equation}
where $\rho_{A/B}$ represent the quantum states associated with sub-systems $A/B$. In this representation, the coefficients $w_{ij}$ denote probabilities, ensuring normalization and non-negativity. Conversely, if the state cannot be decomposed into such separable terms, it is considered quantum mechanically entangled. For pure state, one can verify whether a quantum state is separable or entangled through Schmidt decomposition and von Neumann entropy~\cite{Nielsen:2012yss}. For compound or mixed state, however, it is nontrivial to verify whether it is separable or entangled along with quantifying any measure of entanglement~\cite{Gharibian:2008hgo}. 
Nevertheless, sufficient conditions for quantum entanglement can be obtained through the measure of some entanglement witnesses based on positive maps, linear transformations that map positive matrices to positive matrices~\cite{Horodecki:1996nc}. An implementation of this positive map is a partial transposition map, known as Peres-Horodecki criteria~\cite{Peres:1996dw,Horodecki:1997vt}, which states that a bipartite system is considered entangled if it fails to remain positive under partial transposition. In other words, for a partially transposed density matrix, if there exists at least one negative eigenvalue, the system is said to be entangled. The partial transpose of a density matrix can be done by transposing one of the indices of $A~(B)$ while keeping others fixed. A measure of entanglement is given by the absolute sum of the negative eigenvalues called negativity~\cite{Horodecki:2009zz}. This Peres-Horodecki criterion based on partial transposition is, however, not comprehensive above $2\otimes 3$ level state~\cite{Woronowicz:1975dp}.  

This Peres-Horodecki criterion has been reduced to a simpler form for a system of two qubits through some inequalities relating to the spin correlation parameters of density matrix~\cite{Aguilar-Saavedra:2022uye}. The most general density matrix of a quantum state involving two qubits can be written as~\cite{Bernreuther:1993df,Bernreuther:1993hq},
\begin{equation}
	\label{eqn:fullhalfdm}
	\rho = \frac{1}{4}\left(\mathbb{I}_{4\times 4} + \sum_i(\mathbb{I}_{2\times 2}\otimes p_i^B\sigma_i + p^A_i\sigma_i\otimes\mathbb{I}_{2\times 2}) + \sum_{i,j}c^{AB}_{ij}\sigma_i\otimes\sigma_j\right),
\end{equation}
where $\sigma^i, i\in \{1,2,3\}$ are the Pauli matrices, $p^A_i,p^B_i$ are parameters characterizing spin polarization of qubit $A,$ and $B$ respectively, and $c^{AB}_{ij}$ denotes the spin correlations between the two qubits. The partial transpose~($B$ is transposed) of the density matrix given above can be obtained as,
\begin{equation}
	\label{eqn:partranshalf}
	\rho^T = \frac{1}{4}\left(\mathbb{I}_{4\times 4} + \sum_i \left(\mathbb{I}_{2\times 2}\otimes (p_i^B\sigma_i)^T + p_i^A\sigma_i\otimes\mathbb{I}_{2\times 2}\right) + \sum_{i,j}c_{ij}\left(\sigma_i\otimes\sigma_j^T\right)  \right).
\end{equation}
The measure of entanglement based on the PPT criteria can be estimated as the absolute sum of all the negative eigenvalues $\rho^T$,
\begin{equation}\label{eq:ttqe-ppt}
	{\cal C}^{\rm PPT} = \sum_i\left|\lambda_i^N(\rho^T)\right|,
\end{equation}
where $\lambda_i^N$ are the negative eigenvalues of $\rho^T$. 

Instead of obtaining the eigenvalues of $\rho^T$, the authors in Ref.~\cite{Aguilar-Saavedra:2022uye} suggest to probe the negativity of $v^T\rho^T v$ for different four-vectors $v$. Assuming $v = (1,0,0,\pm1)^T, ~(0,1,\pm1, 0)^T$, the conditions for entanglement~(satisfying any one is sufficient) obtained are,
\begin{equation}
	\label{eqn:corrcond}
	\begin{aligned}
		&|c_{11} + c_{22}| > 1 + c_{33},\\
		&|c_{11} - c_{22}| > 1 -c_{33}.
	\end{aligned}
\end{equation} 
Thus, for the two-qubit case, the measure of entanglement can be done only with the correlation parameters given by
\begin{equation}\label{eq:ttqe-cdiag}
	{\cal C}^\pm = |c_{11} + c_{22}| - 1 \mp c_{33}.
\end{equation}

A different approach for the entanglement signature is also available in literature based on a directly measurable observable~\cite{Afik:2020onf},
\begin{equation}
	D=\frac{{\rm Tr}[c]}{3}=-\frac{1+\delta}{3},
\end{equation}
which can be extracted from the differential cross~section,
\begin{equation}
	\frac{1}{\sigma}\frac{d\sigma}{d\cos\phi_{ll}} = \frac{1}{2}\left(1-D\cos\phi_{ll}\right).
\end{equation} 
Here, $\phi_{ll}$ is the angle between the two charged lepton directions in their respective parent top and anti-top quark's rest frame. The observable $D$ can be obtained from the average value of $\cos\phi_{ll}$ as,
\begin{equation}
	\label{eqn:D}
	D = -3\cdot \langle \cos\phi_{ll}\rangle.
\end{equation}
The existence of entanglement is guaranteed by $D < -1/3$ ($\delta > 0$) and the measure of entanglement is given by
\begin{equation}\label{eq:ttqe-D}
	\mathcal{C}^D\left[\rho\right] = \text{max}\left(0,-1-3D\right)/2.
\end{equation}
ATLAS collaboration has reported the presence of entanglement in $t\bar{t}$ production by measuring the above quantity $D$~\cite{atlascollaboration2023observation}.

In the two qubit system, entanglement is generally quantified through some non-negative function called entanglement monotone~\cite{Horodecki:2009zz}, which does not increase under local operations and classical communication (LOCC). One particular such monotone is concurrence. For two qubit states the concurrence is given by~\cite{Hill:1997pfa,Wootters:1997id},
\begin{equation}\label{eq:ttqe-conc}
	\mathcal{C}\left[\rho\right] = \text{max}\left(0,\lambda_1-\lambda_2-\lambda_3-\lambda_4\right), \quad \lambda_1 \ge \lambda_2 \ge \lambda_3 \ge \lambda_4,
\end{equation}
where $\lambda_i$ are the eigenvalues of the matrix $\mathcal{C}(\rho) = \sqrt{\sqrt{\rho}\widetilde{\rho}\sqrt{\rho}}$, with $\widetilde{\rho}=\left(\sigma_2\otimes\sigma_2\right)\rho^\star\left(\sigma_2\otimes \sigma_2\right)$ is the spin-flipped density matrix and $\rho^\star$ is the complex conjugate of state in Eq.~(\ref{eqn:fullhalfdm}). The concurrence satisfies $0\le \mathcal{C}\left[\rho \right] \le 1$, with a state being entangled if and only if $\mathcal{C}\left[\rho\right]>0$. For maximally entanglement state, $\mathcal{C}\left[\rho\right]=1$.

For systems larger than $2\otimes2$, the above concurrences, however, do not provide reliable answers as they only give an upper bound. For a system larger than $2\otimes2$,  the concurrence is obtained in the following way. For a bipartite pure quantum system comprising two sub-systems $A$ and $B$ of equal dimensionality, the concurrence is defined as~\cite{Rungta_2001},
\begin{equation}
	\mathcal{C}\left[\rho\right] = \sqrt{2\left(1-\text{Tr}\left[(\rho_r)^2\right]\right)}, \quad r \in \{A,B\},
\end{equation}  
where $\rho_r$ is the reduced density matrix obtained by tracing over the degrees of freedom of either sub-system, $\rho_A = \text{Tr}_B\left[\rho\right]$. One can extend the definition of concurrence given above to a general mixed state $\rho$ using the convex roof extension~\cite{Uhlmann:1996mk}, $\mathcal{C}\left[\rho\right] = \text{inf}\sum_i w_i \mathcal{C}\left[\rho_i\right]$ for all possible decomposition into pure states. The construction of $\mathcal{C}\left[\rho\right]$ is accurate only for two-level sub-systems. In contrast, for  $d > 2$, finding $\mathcal{C}\left[\rho\right]$ for any general mixed state via concurrence of the pure states becomes nontrivial. 
Thankfully, there are established lower bounds on the concurrence applicable to a general $m\otimes n (m\le n)$ dimensional mixed quantum state~\cite{PhysRevLett.95.040504,PhysRevLett.95.210501,PhysRevLett.92.167902,PhysRevA.75.052320,PhysRevA.82.044303,PhysRevA.86.054301,PhysRevA.84.052112,PhysRevA.78.042308}. The presence of entanglement can be unmistakably indicated if these bounds are non-zero.  In this article, we use the analytic form for the   bounds presented in Ref.~\cite{PhysRevA.86.054301} for any $m\otimes n(m\le n)$ quantum state $\rho$, where the lower bound is given by,
\begin{equation}\label{eq:em-low-diff-dim}
	\begin{aligned}
		\mathcal{C}\left[\rho\right]_{LB} &= \text{max}\left[0, \sqrt{\frac{2}{m(m-1)}}\left(||\rho^{T_{A/B}}||-1\right),\sqrt{\frac{2}{m(m-1)}}\left(||R(\rho)||-1\right),\right.\\&\left. \sqrt{2\left[\text{Tr}(\rho^2)-\text{Tr}(\rho_A^2)\right]},\sqrt{2\left[\text{Tr}(\rho^2)-\text{Tr}(\rho_B^2)\right]}\right]
	\end{aligned}	
\end{equation}
and the upper bound is given by,
\begin{equation}\label{eq:em-high-diff-dim}
	\mathcal{C}\left[\rho\right]_{UB} = \text{min}\left[\sqrt{2\left[1-\text{Tr}(\rho_A^2)\right]},\sqrt{2\left[1-\text{Tr}(\rho_B^2)\right]}\right].
\end{equation}
The maximum value for $\mathcal{C}\left[\rho\right]$ is obtained for symmetric and maximally entangled pure state. In the above equation, $\rho^{T_{A/B}}$ represents the density matrix obtained by doing a partial transpose of sub-system $A/B$, and $R(\rho)$ is the realigned matrix defined as $R(\rho)_{ij,kl}=\rho_{ik,jl}$, where $i$ and $j$ are the row and column indices for the sub-system $A$ respectively, while $k$ and $l$ are such indices for the sub-system $B$. The $||\cdot||$ denotes the trace norm defined  as $||\rho|| = \text{Tr}\left(\sqrt{\rho^\dagger\rho}\right).$

Another measure of entanglement for bipartite mixed state $\rho$ can be computed based on the trace norm of partial transpose $\rho^{T_A}$. From the trace norm of $\rho^{T_A}$ denoted by $||\rho^{T_A}||$, one can define \emph{negativity}~\cite{PhysRevA.65.032314},
\begin{equation}
	\mathcal{N}(\rho) = \frac{||\rho^{T_A}||-1}{2},
\end{equation}
which corresponds to the absolute value of the sum of the negative eigenvalue of $\rho^{T_A}$, and which vanishes for un-entangled states. Using the trace norm, one can also define another measure of entanglement called \emph{logarithmic negativity}~\cite{PhysRevA.65.032314,PhysRevLett.95.090503},
\begin{equation}\label{eq:logdrho}
	\mathcal{E}_{N}(\rho) = \text{Log}_d\left[||\rho^{T_A}||\right] =  \text{Log}_d\left[\text{Tr}\left(\sqrt{\rho^{{T_A}\dagger}\rho^{T_A}}\right)\right].
\end{equation}
In general, the trace norm of a matrix $A$ is equal to the sum of all the singular values of $A$; when $A$ is hermitian then $||A||$ is equal to the sum of absolute values of all eigenvalues.
The positive value of the logarithmic negativity would ensure the existence of entanglement in a bipartite system~\cite{PhysRevA.65.032314}.  We now discuss the various measures of entanglement when the quantum state is composed of more than two particles; in particular, we discuss tripartite entanglement.

\subsection{Tripartite entanglement}
Greenberger-Horne-Zeilinger~(GZH)~\cite{Greenberger1990-GREBTW-3} observed that the entanglement among more than two particles leads to a conflict with local realism for non-statistical predictions of quantum mechanics. A major difference between quantum entanglement and the classical correlation is that quantum entanglement cannot be freely shared among many parties. For example, given three particles $A, B, C$, and if $A$ is maximally entangled with $B$, then $A$ cannot be simultaneously entangled with $C$. However, a relaxation on this restriction was found in Ref.~\cite{PhysRevA.61.052306} provided entanglement of $A$ with $B$ is not maximal. The three-qubit monogamy discovered by Coffman, Kundu, and Wooters~(CKW)~\cite{PhysRevA.61.052306} can be described as follows; consider $\rho_{ABC}$ to be the state which encodes correlation information between individual particles, and similarly, one can construct the reduced state by tracing out one of the particles. The measure of tripartite entanglement is then quantified as,
\begin{equation}
	\tau_{res} = \mathcal{C}^2_{A(BC)} - \mathcal{C}_{AB}^2 - \mathcal{C}_{AC}^2.
\end{equation} 
In the above equation, $\mathcal{C}_{A(BC)}^2$ is the measure of concurrence for one of the sub-system $A$ with the remaining $BC$ sub-system, and $C^2_{AB}, C^2_{AC}$ are obtained by tracing out one of the qubits.

When dealing with tripartite quantum state, one may write the quantum state $\ket{\psi}$ as totally separable ($\ket{\psi_A}\otimes\ket{\psi_B}\otimes\ket{\psi_C}$) or in terms of bi-separable partitions ($\ket{\psi_i}\otimes\ket{\psi_{jk}}$) or a genuinely entangled state. A multipartite quantum state that is not separable with respect to any bi-partition is known as a genuinely entangled state. Different measures of GME are available in the literature~\cite{article,Meyer_2002,Jin:2022kxb}.

Reference~\cite{article} presents a GME measure based on PPT and CCNR. For a tripartite qudit state $\rho \in H_{123} = H_1^d\otimes H_2^d \otimes H_3^d$, where $d$ is the dimension of each subsystem, if $\text{max}{M(\rho),N(\rho)}\le (1+2d)/3$, then $\rho$ is bipartitely separable; otherwise, it has GME. Here 
\begin{eqnarray}
	M(\rho) &=& \frac{1}{3}\left(||\rho^{T_1}|| + ||\rho^{T_2}||+||\rho^{T_3}||\right),\nonumber\\  
	N(\rho) &=& \frac{1}{3}\left(||R_{1|23}(\rho)|| + ||R_{2|13}(\rho)||+||R_{3|12}(\rho)||\right),	
\end{eqnarray}
where $T_i$s are the partial transpose over the $i^{\text{th}}$ sub-system and $R_{i|jk}$ stands for the bipartite realignment with respect to sub-system $i$ and sub-system $jk$. The measure GME concurrence for a pure state $\ket{\psi}$ is defined by~\cite{article},
\begin{equation}
	\mathcal{C}_{\text{GME}}(\ket{\psi}) = \sqrt{\text{min}\{1-\text{Tr}(\rho_1^2),1-\text{Tr}(\rho_2^2),1-\text{Tr}(\rho_3^2)\}},
\end{equation}
while for the mixed state,
\begin{equation}\label{eq:GME2}
	\mathcal{C}_{\text{GME}}(\rho) \ge \frac{1}{\sqrt{d(d-1)}}\left(\text{max}\{M(\rho),N(\rho)\}-\frac{1+2d}{3}\right).
\end{equation}

Another approach for the measure of GME is available based on the perimeter and area of one-to-other bipartite concurrences.
Considering one-to-other bipartite entanglement, three different concurrences can be obtained, viz. $\mathcal{C}_{1(23)}$ (concurrence between  1 and the sub-system 2 and 3), $\mathcal{C}_{2(31)}$ and $\mathcal{C}_{3(12)}$. These three concurrence   satisfy the monogamy inequality~\cite{PhysRevA.92.062345}, $\mathcal{C}_{i(jk)}^2 \le \mathcal{C}_{j(ki)}^2 + \mathcal{C}_{k(ij)}^2$. 
The perimeter of the triangle formed by the three  ${\mathcal{C}_{i(jk)}}$ is considered to be a tripartite entanglement measure~\cite{Meyer_2002} for three spin-$1/2$ system. An additional measure of GME is obtained for general multipartite states by considering the area of concurrence triangle~\cite{Jin:2022kxb},
\begin{equation}\label{eq:GME-F}
	{\cal F}_{123} \equiv \left[\frac{16}{3}Q(Q-\mathcal{C}_{1(23)})(Q-\mathcal{C}_{2(13)})(Q-\mathcal{C}_{3(12)})\right]^{1/2},
\end{equation} 
where 
\begin{equation}\label{eq:GME-Q}
	Q = \frac{1}{2}\left(\mathcal{C}_{1(23)}+\mathcal{C}_{2(13)}+\mathcal{C}_{3(12)}\right).
\end{equation}
In this article, we employ the measure of GME based on the area of the concurrence triangle to study tripartite entanglement in $2\times2\times3$ quantum states. 
In the next sections, we discuss various measures of entanglement for the production processes $t\bar{t}$, $VV$, and $t\bar{t}Z$ at the LHC  and future $e^+-e^-$ colliders.

	\section{Entanglement  in $t\bar{t}$ pair production}
\label{sec:halfhalf}
The top quark pair  ($t\bar{t}$) production process serves as an exemplary system for investigating QE in the high-energy regime. This process has undergone extensive scrutiny from both theoretical and experimental perspectives to delve into QE with various tests for the presence and measures of QE.  In this article, we present an exhaustive examination of these tests and measures of QE across different phase-spaces at the LHC and future $e^+e^-$ colliders. Our analysis relies on the production density matrix which encapsulates the polarizations of the top quarks and their spin correlations. These polarization and spin-correlations are transferred to their decay products as they decay before hadronization.  By analyzing the angular distributions of the decay products one can measure the polarizations and spin correlations through some asymmetries numerically.  

The production of top-antitop ($t\bar{t}$) pairs at the LHC, is primarily governed by strong interactions. Consequently, these processes exhibit no inherent polarization ($p_i = 0$) at the leading order. This lack of polarization arises due to the conservation of parity~(longitudinal polarization) and the time invariance~(transverse polarization) inherent in quantum chromodynamics~(QCD)~\cite{Mahlon:1995zn}. The less than $1\%$ contribution to both polarization can come from the electroweak and absorptive part at one loop~\cite{Bernreuther:2013aga,Bernreuther:2015yna}. Nevertheless, there is an anticipation of correlated spins for the $t\bar{t}$ pairs, and this spin correlation has already been detected by both the ATLAS and CMS collaborations at the LHC~\cite{ATLAS:2012ao,ATLAS:2014abv,ATLAS:2019zrq,CMS:2013roq,CMS:2019nrx}.

To investigate QE, we use the Monte-Carlo event generator {\tt MadGraph5$\_$aMC$@$NLO}~({\tt MG5} henceforth)~\cite{Alwall:2011uj,Alwall:2014hca}  to simulate $t\bar{t}$ production followed by decaying them fully leptonically ($e$ and $\mu$) providing a final state with two $b$-jet, two oppositely charged leptons and missing energy  at both the LHC and  future $e^+e^-$ colliders.
For simplicity, we restrict our analysis to the parton level without imposing any cuts on the final state particles.  The quantum state~($\rho$) composed of two spin-$\frac{1}{2}$ top quark can be described as in Eq.~(\ref{eqn:fullhalfdm}), where the parameters of density matrix are obtained from the angular distribution of final state leptons. The joint angular distribution of the final state leptons at the rest frame of the top quarks  is given by,
\begin{equation}
	\frac{1}{\sigma}\frac{d^2\sigma}{d\Omega_ld\Omega_{\bar{l}}} = \frac{1}{16\pi^2}\left[1+\alpha_tp_i^tc_i^{t}+\alpha_{\bar{t}}p_i^{\bar{t}}c_i^{\bar{l}} + \alpha_t\alpha_{\bar{t}}c_{ij}^{t\bar{t}}c_i^lc_j^{\bar{l}}\right],
\end{equation}
where $c_i$ are the angular functions associated with the daughter leptons, i.e,
\begin{equation}\label{eq:af-cxcycz}
	c_x = \sin\theta\cos\phi,c_y=\sin\theta\cos\phi,c_z=\cos\theta,
\end{equation}
and $\alpha^\prime$s are the spin analyzing power. The parameters $p^{t/\bar{t}}$, and $c^{t\bar{t}}$ represent the vector polarization of top/anti top quark and vector-vector spin correlation matrix, respectively. The spin analyzing power $\alpha$ for a spin-$1/2$ top quark decaying to one spin-1 $W^+$ boson and spin-$1/2$ $b$-quark through a vertex structure $\bar{f}\gamma^\mu\left(C_LP_L+C_RP_R\right)fV_\mu, P_{L/R} = \frac{1}{2}\left(1\mp \gamma_5\right)$ is given by~\cite{Rahaman:2021fcz},
\begin{equation}
	\alpha_{W^+/b} = \frac{(C_R^2-C_L^2)(1-x_{W^+}^2-2x_b^2)\sqrt{1+(x_{W^+}^2-x_b^2)^2-2(x_{W^+}^2+x_b^2)}}{(C_R^2+C_L^2)(1-2x_{W^+}^2+x_b^2+x_{W^+}^2x_b^2+x_{W^+}^4-2x_b^4)-12C_RC_Lx_{W^+}x_b^2},
\end{equation}
where $x_i=m_i/m_t$ with $m_i$ the mass of the daughters and $m_t$ as the mass of the top quark. At the leading order, for top quark decay $t\to bW^+$, we have $\alpha_{W^+/b} \approx -0.396$ within SM. One has to note that in the collider experiments, the accurate reconstruction of angular differential functions becomes non-trivial due to two undetected neutrinos. In this article, we do not perform reconstruction of neutrinos which can be otherwise achieved with various reconstruction techniques like neutrino weighting method~\cite{ATLAS:2019zrq}, $M_{T2}$ assisted on-shell~(MAOS) algorithm~\cite{Barr:2003rg,Lester:1999tx,Cho:2008tj,Rahaman:2023pte}, or using machine learning techniques~\cite{Raine:2023fko}.

The polarization and correlation parameters are obtained from the asymmetries in angular functions as~\cite{Rahaman:2021fcz},
\begin{equation}
	\label{eqn:halfasymm}
	\begin{aligned}
		p_i^{t/\bar{t}} &= \frac{2}{\alpha_{t/\bar{t}}}\frac{\sigma(c_i > 0) - \sigma(c_i <0)}{\sigma(c_i > 0) + \sigma(c_i < 0)},\\
		c_{ij}^{t\bar{t}} &=  \frac{4}{\alpha_t\alpha_{\bar{t}}}\frac{\sigma(c_ic_j > 0) - \sigma(c_ic_j <0)}{\sigma(c_ic_j > 0) + \sigma(c_ic_j < 0)}.
	\end{aligned}	
\end{equation}
Any error in these asymmetries will propagate to the elements of the density matrix and finally to the measure of QE. We estimate Monte-Carlo error to the measure in QE as follows. We randomly select $20\%$ of events from a total of {\tt 1M} events 1000 times and for each subset the corresponding asymmetries are calculated followed by the measure of QE. The deviation in the measure of QE is obtained from the Gaussian fit. We report $1\sigma$ MC deviation in our measurement. 

The reference frame in which the angles are measured is defined as follows: the $z$-axis aligns with the boost direction of the visible particles. This alignment occurs because the boost typically follows the direction of the colliding quark, owing to the higher probability of quarks than anti-quarks inside the proton. The $xz$ plane, from which azimuth angles ($\phi$) are measured, is defined by the production plane—specifically, the plane formed by the $z$-axis and the direction of the top quark.
The polar angles $\theta$ and azimuth angles $\phi$ of the final leptons are constructed in the center-of-mass frame of the $t\bar{t}$ system, and the top quark pairs are Lorentz boosted to this rest frame.

\begin{table}[!h]
	\centering
	\caption{\label{tab:ppttbar}
		Different measures of entanglement estimated with the mean number of events in various bins of invariant mass for the $t\bar{t}$ are listed for $t\bar{t}$ production at LHC with $\sqrt{s}=13$~TeV. 
		The error corresponds to the 1$\sigma$ MC error, with estimation methods described in the text. }
	\renewcommand{\arraystretch}{1.5}
	\begin{tabular*}{\textwidth}{@{\extracolsep{\fill}}lcccccc@{}}\hline	
		Bins & Events & ${\cal C}^{+}$ &${\cal C}^-$& ${\cal C}^{\text{PPT}}$ & $(-1-3D)/2$ & ${\cal C}[\rho]$ \\ \hline
		Unbin & $200000$ & $0.523\pm 0.013$ &$-1.075\pm0.013$& $0.506\pm0.004$ & $-0.158\pm 0.005$ & $-0.245\pm 0.005$ \\
		M$_1$ & $40393$ & $0.223\pm 0.022$ &$-0.714\pm0.023$& $0.275\pm0.005$ & $0.121\pm0.016$ & $-0.352\pm 0.015$ \\
		M$_2$ & $75203$ & $0.641\pm 0.024$ &$-1.142\pm0.024$& $0.570\pm0.007$ & $-0.105\pm 0.009$& $-0.237\pm0.009$ \\
		M$_3$ & $39994$ & $0.766\pm 0.032$ &$-1.372\pm0.031$& $0.727\pm0.010$ & $-0.301\pm0.011$ & $-0.377\pm0.019$ \\
		M$_4$ & $20129$ & $0.806\pm 0.045$ &$-1.478\pm0.045$& $0.791\pm0.014$ & $-0.403\pm0.016$ & $-0.430\pm0.025$ \\
		M$_5$ & $10408$ & $0.789\pm 0.060$ &$-1.479\pm0.061$& $0.792\pm0.019$ & $-0.465\pm0.023$ & $-0.393\pm0.036$ \\
		M$_6$ & $5620$ & $0.645\pm0.076$ &$-1.441\pm0.084$& $0.774\pm0.024$ & $-0.518\pm0.031$ & $-0.343\pm0.053$ \\
		M$_7$ & $3169$ & $0.595\pm0.091$ &$-1.401\pm0.111$& $0.762\pm0.032$ & $-0.495\pm0.040$ & $-0.321\pm0.078$ \\
		M$_8$ & $5079$ & $0.470\pm0.067$ &$-1.200\pm0.083$& $0.742\pm0.017$ & $-0.630\pm0.032$ & $0.208\pm0.140$ \\
		\hline		
	\end{tabular*}
\end{table}
Here, we measure QE using four different approaches as described in the earlier section given in Eqs.~(\ref{eq:ttqe-ppt}), (\ref{eq:ttqe-cdiag}), (\ref{eq:ttqe-D}), and (\ref{eq:ttqe-conc}).  For the LHC, we estimate the measures of QE in selected regions of the invariant mass of the top quarks ($m_{t\bar{t}}$) including the full phase-space and they are listed in Table~\ref{tab:ppttbar} with 1$\sigma$ MC error, with estimation methods described above. The regions or bins in $m_{t\bar{t}}$ are chosen as
\begin{eqnarray}\label{eq:mtt-bins}
	\text{M}_1&\equiv&m_{t\bar{t}}\le 400~\text{GeV},\nonumber\\
	\text{M}_{n+1} &\equiv& 400+100\times(n-1) < m_{t\bar{t}}\le 400+100\times n~\text{GeV},~n=1,2\ldots6,\nonumber \\
	\text{M}_8&\equiv& m_{t\bar{t}} > 1000~\text{GeV}.
\end{eqnarray}
The mean number of events randomly drawn from a total of 1M events (20\%) are listed in the second column corresponding to the bins in the first column. The measures of QE based on diagonal spin correlations (${\cal C}^\pm$), negativity of the PPT criterion ($\cal {C}^{\rm PPT}$), $D$ marker based on angular separation of two leptons at the top and anti-tops rest frame ($-(1+3D)/2$), and the concurrence with spin-flipped $\rho$ (${\cal C}[\rho]$) are listed in the third, fourth, fifth, sixth, and last columns, respectively. The measure ${\cal C}^+ > 0$ provides a sufficient criterion for the presence of QE in unbinned and all bins of $m_{t\bar{t}}$, although ${\cal C}^- < 0$. The $\cal {C}^{\rm PPT}$ measure also indicates the presence of QE in all $m_{t\bar{t}}$ regions.  

However, the $D$ marker QE measure and the concurrence ${\cal C}[\rho]$ suggest that the $t\bar{t}$ system is not entangled in all regions of $m_{t\bar{t}}$. The condition for entangled $t\bar{t}$ based on the $D$ marker, i.e., $D<-1/3$ or $(-1-3D)/2 > 0$, indicates that the $t\bar{t}$ state is entangled only at $m_{t\bar{t}} < 400$~GeV (threshold region), while it is separable for all other regions of $m_{t\bar{t}}$. Conversely, the concurrence ${\cal C}[\rho]$ indicates the presence of QE in a large region of the invariant mass, $m_{t\bar{t}}>1000$~GeV, which is the region mostly dominated by the $q\bar{q}$ channel, producing a mixed $t\bar{t}$ state.

The $t\bar{t}$ pairs at LHC are dominantly produced via. gluon-gluon fusion. When the $t\bar{t}$ pairs are produced close to the threshold, approximately $80\%$ of the production cross~section arises from a spin-singlet state~\cite{Afik:2020onf,Kiyo:2008bv,Kuhn:1992qw,Petrelli:1997ge}, which is maximally entangled. After averaging over all possible top-quark
directions, entanglement only survives at the threshold because of the rotational invariance of the spin-singlet. The behavior has been explained in detail in Ref.~\cite{Afik:2020onf}, where the authors show how $t\bar{t}$ pairs produced from $gg$-fusion are entangled at the threshold and large $m_{t\bar{t}}$ region.

To compare with the ATLAS result of quantum entanglement (QE), we also measure the $D$ marker in the threshold region $340 < m_{t\bar{t}}\le 380$~GeV, yielding $D = -0.534 \pm 0.010$  in contrast with the  ATLAS~\cite{atlascollaboration2023observation}  measure of $D=-0.547\pm0.002 (\text{stat.})\pm0.021 (\text{syst.})$. It's worth noting that our estimate is much more simplistic, as our analysis is conducted solely at the parton level, utilizes truth information of neutrino momenta, and does not account for any background effects."

\begin{table}[!h]
	\centering
	\caption{\label{tab:eettbar}		
		Different measures of entanglement estimated with the mean number of events in various bins of invariant mass for the $t\bar{t}$ are listed for $t\bar{t}$ production at $e^+-e^-$ collider with $\sqrt{s}=500$~GeV. 
		The error corresponds to the 1$\sigma$ MC error, with estimation methods described in the text.}
	\renewcommand{\arraystretch}{1.5}
	\begin{tabular*}{\textwidth}{@{\extracolsep{\fill}}lcccccc@{}}\hline	
		Bins & Events & ${\cal C}^{+}$ &${\cal C}^-$& ${\cal C}^{\text{PPT}}$ & $(-1-3D)/2$ & ${\cal C}[\rho]$ \\ \hline	
		Unbin & $200000$ & $0.260\pm 0.014$ &$-1.292\pm0.014$& $0.065\pm0.003$ & $-0.995\pm0.005$ & $0.128 \pm 0.007$ \\
		C$_1$ & $10775$ & $0.195\pm 0.059$ &$-1.457\pm0.060$& $0.078 \pm 0.016$ & $-0.992\pm0.010$ & $-0.077\pm 0.034$ \\
		C$_2$ & $12520$ & $0.511\pm0.055$ &$-0.745\pm0.056$& $0.131\pm0.014$ & $-0.993\pm0.011$ & $0.106\pm 0.034$ \\
		C$_3$ & $15560$ & $0.567\pm0.048$ &$-0.585\pm0.050$& $0.143\pm0.012$ & $-1.0\pm0.013$ & $0.261\pm 0.024$ \\
		C$_4$ & $19592$ & $0.483\pm0.042$ &$-0.690\pm0.044$& $0.122\pm0.011$ & $-1.006\pm0.014$ & $0.189\pm 0.024$ \\
		C$_5$ & $24915$ & $0.398\pm0.039$ &$-0.982\pm0.039$& $0.100\pm0.010$ & $-0.998\pm0.016$ & $0.053 \pm 0.025$ \\
		C$_6$ & $31086$ & $0.257\pm0.036$ &$-1.256\pm0.036$& $0.071\pm0.009$ & $-0.986\pm0.019$ & $-0.011 \pm 0.021$ \\
		C$_7$ & $38476$ & $0.139\pm0.030$ &$-1.570\pm0.032$& $0.045\pm0.008$ & $-0.985\pm0.021$ & $-0.025 \pm 0.017$ \\
		C$_8$ & $47071$ & $0.041\pm0.028$ &$-1.846\pm0.029$& $0.017\pm0.007$ & $-0.996\pm0.022$ & $0.003 \pm 0.013$ \\
		\hline		
	\end{tabular*}
\end{table}
We perform a similar measurement for the QE in $t\bar{t}$ pair production at the future $e^+-e^-$ collider where one can exploit the potential of using initially polarized beams. For this article, we reserved the use of polarized beams for the analysis of the qutrit-qutrit case with di-boson processes in the next section. We simulate the $t\bar{t}$ process  at the center of mass energy of $500$~GeV of the $e^+e^-$ beams in {\tt MG5} with unpolarized beams. In this case, we binned the events in the region of $\cos\theta_t$, where $\theta_t$ is the production angle of the top quark measured with respect to $e^-$ beam. The region in $\cos\theta_t$ are chosen as follows:
\begin{equation}
	{\rm C}_n \equiv -1+\frac{(n-1)}{4} <\cos\theta_t\le -1+\frac{n}{4},~ n=1,2\ldots8.
\end{equation}
The measure of QE  is performed with four different approaches as was  done for the LHC case   and they are listed in Table~\ref{tab:eettbar} along with the mean number of events used for the estimate. 	 

The measures ${\cal C}^+$ and ${\cal C}^{\rm PPT}$ both suggest the presence of QE in all bins of $\cos\theta_t$, while the $D$ marker indicates the absence of QE in all $\cos\theta_t$ region. The concurrence measure, on the other hand, indicates an entangled $t\bar{t}$ state except for C$_1$, C$_6$, C$_7$, and C$_8$ regions of production angle.

\begin{figure}[!htb]
	\centering
	\includegraphics[width=0.4\textwidth]{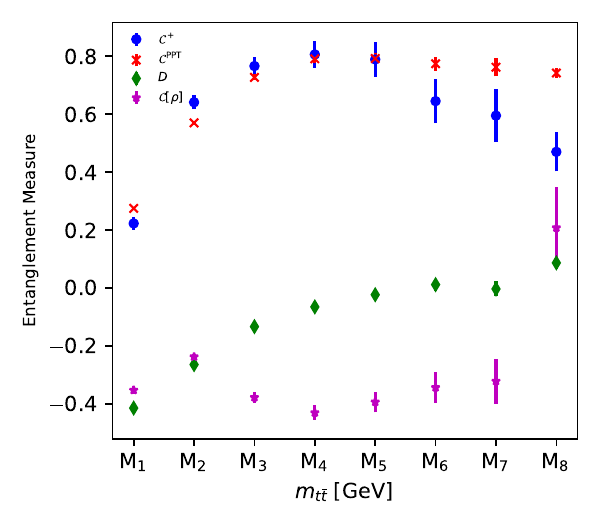}
	\includegraphics[width=0.4\textwidth]{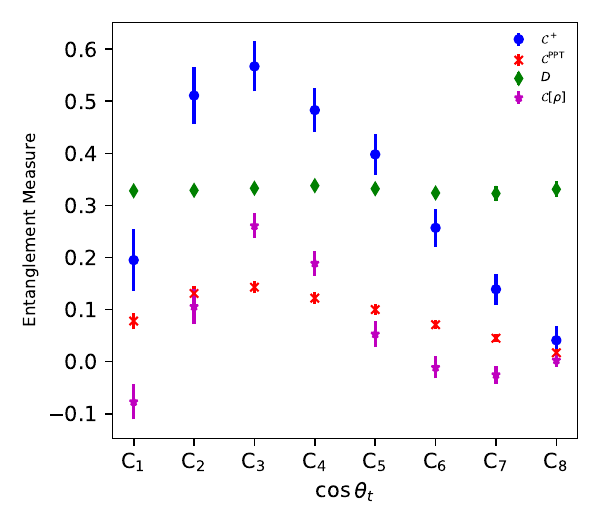}
	\caption{\label{fig:ttbar}
		Various measures of QE are shown for $t\bar{t}$ pair production process at the LHC ({\em left-panel}) and future $e^+-e^-$ collider ({\em right-panel}) at 	$\sqrt{s}=13$~TeV and 	$500$~GeV, respectively.  Among ${\cal C}^\pm$, only ${\cal C}^+$ is shown as anyone being positive  is a sufficient condition for the presence of entanglement (${\cal C}^- <0$ always ). The ${\cal C}^+$ measure is represented by blue points, the PPT measure by red points, the QE measure based on $D$ by green points, and the concurrence by purple points. The error bars with points correspond to 1$\sigma$ Monte Carlo errors in the measurements. See texts for details. }
\end{figure}	 
For graphical representation, we show different measures of QE in Fig.~\ref{fig:ttbar}
for the LHC in {\em left-panel} and for $e+-e^-$ collider in {\em right-panel}. Among ${\cal C}^\pm$, only ${\cal C}^+$ is shown as anyone being positive  is a sufficient condition for the presence of entanglement, ${\cal C}^- <0$ in both collider in all bins considered here. 

We conclude this section by noting that top quark pair production in high-energy colliders provides a relativistic platform to explore quantum entanglement. The extremely short lifetime of top quarks enables the determination of their polarizations and spin correlations, facilitating the reconstruction of the full density matrix necessary for measuring QE. 
We observed discrepancies among different measures of QE, including those derived from the diagonal elements of the correlation matrix, the positive partial transpose (PPT) criterion, the $D$-value, and concurrence, across various phase-space bins. The ${\cal C}^+$ and ${\cal C}^{\rm PPT}$ measures suggest the presence of QE in $t\bar{t}$ production across all phase-space regions at both the LHC and $e^+-e^-$ collider. However, the $D$ value and concurrence highlight the presence of QE only at the threshold and high invariant mass regions, respectively, at the LHC. 
The entanglement around the threshold region at the LHC can be understood by the dominant production of spin-singlet $t\bar{t}$ states via gluon-gluon fusion. Conversely, in regions of large invariant mass, $t\bar{t}$ exists in a spin-triplet state, which is also a maximally entangled state, as emphasized by concurrence. 
In the case of the $e^+-e^-$ collider, the $D$ marker does not detect the presence of QE in any $\cos\theta_t$ region, while the concurrence measure highlights the presence of QE in a few $\cos\theta_t$ regions.

	\section{Entanglement  in $tW^-$ production}
\label{sec:halfone}	
The $tW^-$ production process offers a unique opportunity to explore the entanglement of a less explored $2\otimes 3$ quantum state. Various entanglement measures used for the two-qubit case are not directly applicable in this system due to the presence of one qutrit ($W$). However, one can still use the PPT criterion to check for an entangled state and measure QE. Additionally, we use the lower bound presented in Eq.~(\ref{eq:em-low-diff-dim}) for a general $m\otimes n$ dimension to explore QE in the $tW^-$ process. To investigate QE in the $tW^-$ production process, we simulate the production followed by the leptonic ($e$, $\mu$) decay of the top quark and $W^-$ boson, leading to a $2\ell+ b+\cancel{E}_T$ final state, in {\tt MG5} for the LHC with $\sqrt{s}=13$~TeV. At the leading order, $tW^-$ production in the SM occurs through an $s$-channel mediated by the $b$ quark and a $t$-channel mediated by the top quark. Similar to the $t\bar{t}$ case in the previous section, we reconstruct the density matrix parameterized by the polarization and spin correlation of the top quark and $W^-$ boson from the angular distribution of the decayed leptons. The reconstructed density matrix is then used for various tests of entanglement in the $tW^-$ production process.
The full spin correlated  density matrix for a system of spin-$\frac{1}{2}$ top quark and spin-$1$ $W^-$ boson can be written as~\cite{Rahaman:2021fcz},
\begin{eqnarray}\label{eq:half-one}
	\rho^{tW^-}
	&=& \frac{1}{6}\Big[ \mathbb{I}_{6\times 6}  + \vec{p}^t\cdot\vec{\sigma}\otimes \mathbb{I}_{3\times 3} + \dfrac{3}{2}\mathbb{I}_{2\times 2} \otimes \vec{p}^{W^-}\cdot\vec{S} 
	+\sqrt{\frac{3}{2}}  \mathbb{I}_{2\times 2} \otimes T_{ij}^{W^-}\big(S_iS_j+S_jS_i\big)\nonumber\\
	&+&\frac{3}{2} pp_{ij}^{tW^-}\sigma_i\otimes S_j + \sqrt{\frac{3}{2}} pT_{ijk}^{tW^-}\sigma_i\otimes (S_jS_k+S_kS_j) \Big].
\end{eqnarray} 
where $p^{t/W^-},T^{W^-}$ are the vector and tensorial polarization parameters of particle~($t$/$W^-$), and $pp/pT^{tW^-}$ denotes vector-vector/vector-tensorial correlation parameters of top quark and $W^-$ boson. The vector $\vec{S}\in \{S_x,S_y,S_z\}$ are three spin operators defined as,
\begin{equation}
	S_x = \frac{1}{\sqrt{2}}\begin{pmatrix}
		0&1&0\\1&0&1\\0&1&0
	\end{pmatrix}, \quad S_y = \frac{1}{\sqrt{2}}\begin{pmatrix}
		0&-i&0\\i&0&i\\0&i&0
	\end{pmatrix},\quad S_z = \begin{pmatrix}
		1&0&0\\0&0&0\\0&0&-1
	\end{pmatrix}.
\end{equation} 
We have introduced extra factors of $\sqrt{3/2}$ and $3/2$ in the last and second-last term in Eq.~(\ref{eq:half-one}) against Ref.\cite{Rahaman:2021fcz} to make the $tW^-$ state separable when correlations are direct product of the polarization of two particles, i.e., $\rho_{tW^-}=\rho_t\otimes\rho_{W^-}$ if $pp^{tW^-}=p^t\times p^{W^-}$ and $pT^{tW^-}=p^t\times T^{W^-}$.

The parameters of the polarization density matrix are constructed from the asymmetries of angular distribution of final leptons in the rest frame of the top quark and $W^-$ boson in the center of mass frame of $tW^-$ system. The asymmetries for vector and tensorial polarizations related to the $W^-$ boson are given by~\cite{Rahaman:2016pqj,Rahaman:2021fcz},
\begin{eqnarray}
	\label{eqn:tw1}
	\mathcal{A}[p^{W^-}_i] &=& \frac{\sigma(c_i^{l^-}>0)-\sigma(c_i^{l^-}<0)}{\sigma(c_i^{l^-}>0)+\sigma(c_i^{l^-}<0)}
	= \frac{3}{4}\alpha_{l^-}p_i^{W^-}, i \in \{x,y,z\},\nonumber\\
	\mathcal{A}[T_{ij}^{W^-}] &=& \frac{\sigma(c_i^{l^-}c_j^{l^-}>0)-\sigma(c_i^{l^-}c_j^{l^-}<0)}{\sigma(c_i^{l^-}c_j^{l^-}>0)+\sigma(c_i^{l^-}c_j^{l^-}<0)} =\frac{2}{\pi}\sqrt{\frac{2}{3}}(1-3\delta_{l^-})T^{W^-}_{ij}~ (i \neq j),\nonumber\\ 
	\mathcal{A}[T_{11-22}^{W^-}] &=& \frac{\sigma((c_1^{l^-}c_1^{l^-}-c_2^{l^-}c_2^{l^-})>0)-\sigma((c_1^{l^-}c_1^{l^-}-c_2^{l^-}c_2^{l^-})<0)}{\sigma((c_1^{l^-}c_1^{l^-}-c_2^{l^-}c_2^{l^-})>0)+\sigma((c_1^{l^-}c_1^{l^-}-c_2^{l^-}c_2^{l^-})<0)}=\frac{1}{\pi}\sqrt{\frac{2}{3}}(1-3\delta_{l^-})T^{W^-}_{11-22},\nonumber\\
	\mathcal{A}[T^{W^-}_{33}] &=& \frac{\sigma(c_3^{l^-}c_3^{l^-}>0)-\sigma(c_3^{l^-}c_3^{l^-}<0)}{\sigma(c_3^{l^-}c_3^{l^-}>0)+\sigma(c_3^{l^-}c_3^{l^-}<0)}= \frac{3}{8}\sqrt{\frac{3}{2}}(1-3\delta_{l^-})T^{W^-}_{33}.
\end{eqnarray} 
The vector-vector and independent vector-tensor spin correlations asymmetries are given by~\cite{Rahaman:2021fcz},
\begin{eqnarray}
	\label{eqn:tw2}
	\mathcal{A}[pp^{tW^-}_{ij}] &=& \frac{\sigma(c_i^{l^+}c_j^{l^-}>0)-\sigma(c_i^{l^+}c_j^{l^-}<0)}{\sigma(c_i^{l^+}c_j^{l^-}>0)+\sigma(c_i^{l^+}c_j^{l^-}<0)}
	=\frac{3}{8}\alpha_{l^+}\alpha_{l^-}pp^{tW^-}_{ij},\nonumber\\
	\mathcal{A}[pT^{tW^-}_{i(jk)}] &=& \frac{\sigma(c_i^{l^+}c_j^{l^-}c_k^{l^-}>0)-\sigma(c_i^{l^+}c_j^{l^-}c_k^{l^-}<0)}{\sigma(c_i^{l^+}c_j^{l^-}c_k^{l^-}>0)+\sigma(c_i^{l^+}c_j^{l^-}c_k^{l^-}<0)}
	=\sqrt{\frac{3}{2}}\frac{2}{3\pi}\alpha_{l^+}(1-3\delta_{l^-})pT^{tW^-}_{i(jk)},\nonumber\\
	\mathcal{A}[pT^{tW^-}_{i(11-22)}] &=& \frac{\sigma(c_i^{l^+}((c_1^{l^-})^2-(c_2^{l^-})^2)>0)-\sigma(c_i^{l^+}((c_1^{l^-})^2-(c_2^{l^-})^2)<0)}{\sigma(c_i^{l^+}((c_1^{l^-})^2-(c_2^{l^-})^2)>0)+\sigma(c_i^{l^+}((c_1^{l^-})^2-(c_2^{l^-})^2)<0)}
	=\sqrt{\frac{3}{2}}\frac{1}{3\pi}\alpha_{l^+}(1-3\delta_{l^-})pT^{tW^-}_{i(11-22)},\nonumber\\
	\mathcal{A}[pT^{tW^-}_{i(33)}] &=& \frac{\sigma(c_i^{l^+}\sin(3\theta^{l^-}) > 0)-\sigma(c_i^{l^+}\sin(3\theta^{l^-})<0)}{\sigma(c_i^{l^+}\sin(3\theta^{l^-}) > 0)+\sigma(c_i^{l^+}\sin(3\theta^{l^-})<0)}
	=\sqrt{\frac{3}{2}}\frac{3}{16}\alpha_{l^+}(1-3\delta_{l^-})pT^{tW^-}_{i(33)}.
\end{eqnarray}
Here,  $c_i$ are the angular functions of the leptons given in Eq.~(\ref{eq:af-cxcycz}).
For the case of $W^-\to l^-\bar{\nu}_l$ decay to two fermions with $V-A$ interaction, the parameters $\alpha_{l^-}$ and $\delta_{l^-}$ are given by~\cite{Boudjema:2009fz},
\begin{equation}
	\label{eqn:walpha}
	\begin{aligned}
		\alpha_{l^-} &= \frac{2(C_R^2-C_L^2)\sqrt{1+(x_{l^-}^2-x_{\nu}^2)^2-2(x_{l^-}^2+x_{\nu}^2)}}{12C_RC_Rx_{l^-}x_\nu+(C_R^2+C_L^2)\left[2-(x_{l^-}^2-x_\nu^2)^2+(x_{l^-}^2+x_\nu^2)\right]},\\
		\delta_{l^-} &= \frac{4C_RC_Lx_{l^-}x_\nu+(C_R^2+C_L^2)\left[x_{l^-}^2+x_\nu^2-(x_{l^-}^2-x_\nu^2)^2\right]}{12C_RC_Lx_{l^-}x_\nu+(C_R^2+C_L^2)\left[2-(x_{l^-}^2-x_\nu^2)^2+(x_{l^-}^2+x_\nu^2)\right]}.
	\end{aligned}	
\end{equation} 
In a high energy limit, the final state fermions are practically massless, i.e., $x_i\to  0,\delta \to 0$, and $\alpha_{l^-} \to (C_R^2-C_L^2)/(C_R^2+C_L^2)$. For the decay of $W^-$ boson, within SM $C_R=0$; thus $\alpha_{l^-} = -1$.

We calculate the relevant asymmetries using the angular functions related to the final decayed leptons at the parton level as in Eq.~(\ref{eqn:tw1})-(\ref{eqn:tw2}). A realistic analysis requires the reconstruction of the 4-momenta of two neutrinos, which becomes non-trivial given only 6 independent kinematic constraints are available. Several techniques like MAOS~\cite{Rahaman:2023pte} and ML algorithms~\cite{Raine:2023fko} can be used to reconstruct the two neutrinos in the $tW$ process. Once the asymmetries are calculated, the parameters of the density matrix in Eqn.~(\ref{eq:half-one}) are obtained using those asymmetries.

\begin{table}[!b]
	\caption{\label{tab:entho}
		PPT entanglement, lower bound, and upper bound of concurrence are listed for the $tW^-$ production process in different bins of $m_{tW^-}$ and the full phase-space at the $13$ TeV LHC. The average number of events used to estimate these measures is also listed in the second column. The error corresponds to the 1$\sigma$ MC error.}
	\renewcommand{\arraystretch}{1.5}
	\begin{tabular*}{\textwidth}{@{\extracolsep{\fill}}lcccc@{}}\hline	
		Bins&Events&${\cal C}^{\rm PPT}$&Lower Bound (${\cal C}[\rho]_{\rm LB}$) &Upper Bound (${\cal C}[\rho]_{\rm UB}$)\\\hline
		Unbin&$200000$&$0$&$0.013\pm 0.004$&$0.966\pm0.001$\\
		M$_1\equiv m_{tW^-}<300$&$48129$&$0$&$0.030\pm0.007$&$0.977\pm0.002$\\
		M$_2\equiv 300\le m_{tW^-}<400$&$84627$&$0$&$0.014\pm0.006$&$0.956\pm0.002$\\
		M$_3\equiv 400\le m_{tW^-}<500$&$36311$&$0$&$0.031\pm 0.009$&$0.960\pm0.003$\\
		M$_4\equiv 500\le m_{tW^-}<600$&$15624$&$0.011\pm0.009$&$0.050\pm 0.015$&$0.966\pm0.004$\\
		M$_5\equiv m_{tW^-}>500$&$15312$&$0.012\pm0.009$&$0.046\pm 0.016$&$0.985\pm0.003$\\	
		\hline
	\end{tabular*}
\end{table}
\begin{figure}[!h]
	\centering
	\includegraphics[width=0.32\textwidth]{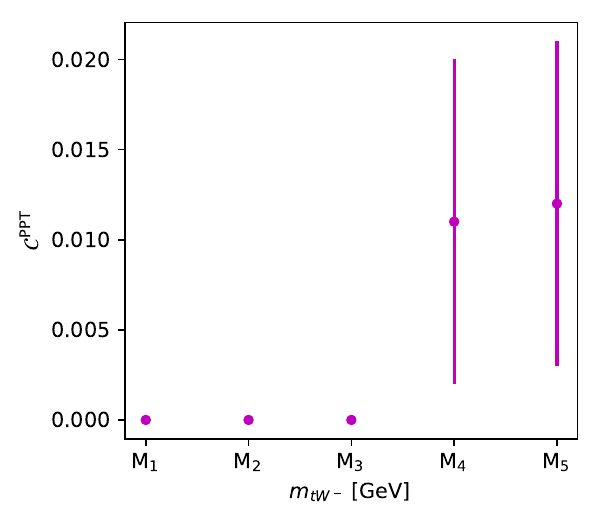}
	\includegraphics[width=0.32\textwidth]{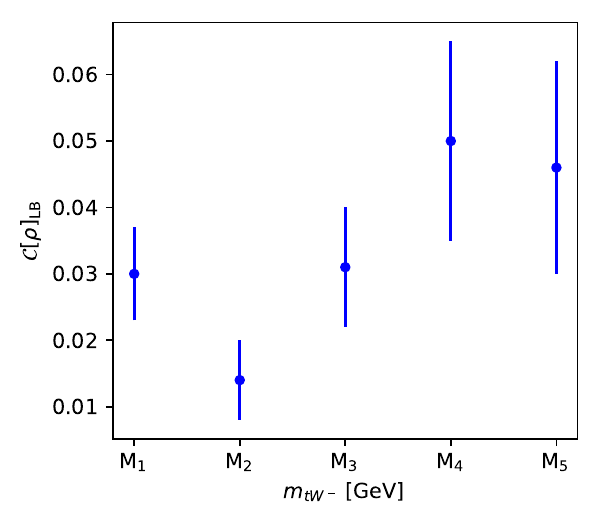}
	\includegraphics[width=0.32\textwidth]{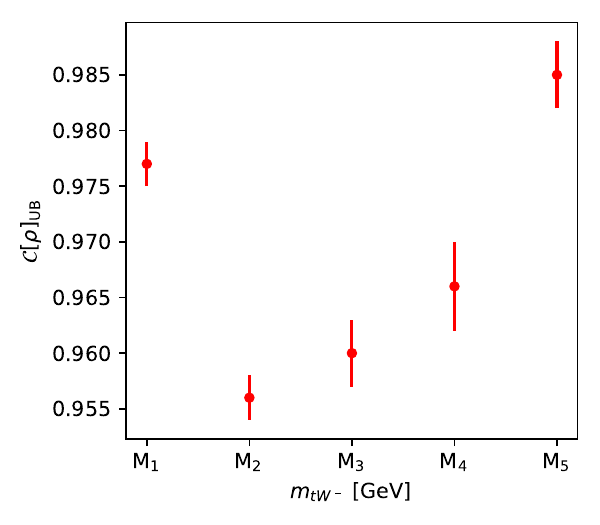}
	\caption{\label{fig:lbubtw}Diagrammatic representation  of entanglement measure based on PPT criterion ({\em left-panel}), lower bound ({\em middle-panel}) and upper bound ({\em right panel}) on concurrence in $tW^-$ production process at the $13$ TeV LHC for different bins of $tW^-$ invariant mass. The error bars correspond to 1$\sigma$ MC error.}
\end{figure} 
Finally, we estimate the entanglement measures in $tW^-$ using the reconstructed density matrix. First, we use the PPT criterion (Eq.~(\ref{eq:ttqe-ppt})) for checking the presence of entanglement and estimating its measure. We then calculate the lower and upper bounds of entanglement as given in Eqs.~(\ref{eq:em-low-diff-dim}) and (\ref{eq:em-high-diff-dim}), respectively. The MC error on the entanglement measure is estimated by randomly selecting $20\%$ of the total {\tt 1M} events and calculating all measures, similar to the procedure used for the $t\bar{t}$ case in the previous section. We calculate the entanglement measures in a few chosen regions of $tW^-$ invariant mass, including the full phase-space, and they are listed in Table~\ref{tab:entho} along with the 1$\sigma$ MC errors. In the same table, we also list the mean number of events used to calculate the entanglement measures for each bin in the second column.

The PPT entanglement, ${\cal C}^{\rm PPT}$, is positive only for $500\le m_{tW^-}<600$ and $m_{tW^-}>500$, while zero for the other bins and unbin region of invariant mass. 
However, the positive ${\cal C}^{\rm PPT}$ values are consistent with zero above the 2$\sigma$ MC error. On the other hand, the lower bound of concurrence, ${\cal C}[\rho]_{\rm LB}$, indicates the presence of entanglement in the $tW^-$ state in all region of invariant mass within 2$\sigma$ MC error.  The upper bounds on the concurrence are all close to $1$ as they should be. For pictorial presentation,  these measures of entanglement are shown in Fig.~\ref{fig:lbubtw}.

	\section{Entanglement in vector boson pair production}
\label{sec:oneone}
\begin{figure}[!htb]
	\centering
	\includegraphics[width=0.32\textwidth]{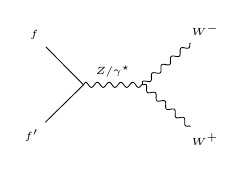}
	\includegraphics[width=0.32\textwidth]{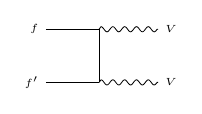}
	\includegraphics[width=0.32\textwidth]{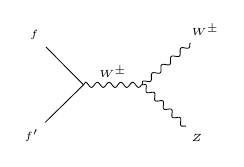}
	\caption{\label{fig:vv}Schematic Feynman diagrams at the leading order for the di-boson production processes. The $VV$ in $t$-channel represents $W^-W^+/WZ/ZZ$ process. }
\end{figure}
Vector boson pair production, such as $W^+W^-$, $W^\pm Z$, and $ZZ$ pair production, has recently gained a lot of interest for the exploration of quantum entanglement and violation of Bell's inequality, as these processes offer a unique ground for $3\times 3$ quantum states at high energy. While not intending to summarize results available in the literature, we provide measures of entanglement in these vector boson pair production processes ($W^+W^-$, $W^+ Z$, and $ZZ$) using two different approaches and compare them from the density matrices reconstructed from the measurement of polarization and spin correlation. At the leading order in the SM, the Feynman diagrams representing $W^+W^-$, $W^+Z$, and $ZZ$ di-boson production processes are shown in Fig.~\ref{fig:vv}. We simulate these $VV$ ($V=W,Z$) processes in {\tt MG5} in the fully leptonic ($e$, $\mu$) final state and use the parton level events with true neutrino momenta to measure the polarization and spin correlation of the massive vector bosons. The missing neutrinos can, in principle, be reconstructed using MAOS and machine learning approaches with reasonable reconstruction accuracy.
The full spin correlated density matrix for two spin-$1$ boson~$(A,B)$ is given by~\cite{Rahaman:2021fcz},
\begin{equation}
	\label{eqn:spin1spin1}
	\begin{aligned}
		\rho_{AB} &= \frac{1}{9}\left[\mathbb{I}_{9\times 9}+\frac{3}{2}\vec{p}^{A}\cdot \vec{S}\otimes \mathbb{I}_{3\times 3} + \frac{3}{2}\mathbb{I}_{3\times 3}\otimes \vec{p}^{B}\cdot\vec{S} + \sqrt{\frac{3}{2}}T_{ij}^{A}(S_iS_j+S_jS_i)\otimes \mathbb{I}_{3\times 3}\right.\\&+\left. \sqrt{\frac{3}{2}}\mathbb{I}_{3\times 3}T^{B}_{ij}(S_iS_j+S_jS_i) + \frac{9}{4}pp^{AB}_{ij}S_i\otimes S_j + \frac{3}{2}\sqrt{\frac{3}{2}}pT^{AB}_{ijk}S_i\otimes(S_jS_k+S_kS_j)\right.\\&+\left.\frac{3}{2}\sqrt{\frac{3}{2}}Tp^{AB}_{ijkj}(S_iS_j+S_jS_i)\otimes S_k + \frac{3}{2}TT^{AB}_{ijkl}(S_iS_j+S_jS_i)\otimes(S_kS_l+S_lS_k)\right].
	\end{aligned}
\end{equation}
Here $p^{A/B}$, and $T^{A/B}$ denotes the vector and tensor polarizations of boson $A/B$, and $pp^{AB},pT^{AB},TT^{AB}$ represents vector-vector, vector-tensor, and tensor-tensor correlations, respectively of two bosons. The $Tp^{AB}_{ijk}$ are identical to $pT^{BA}_{kij}$. We introduced an extra coefficient of $3/2$ in the spin correlation $pp$ and $TT$,  and  $\sqrt{3/2}$ in the spin-correlations $pT$ against Ref.\cite{Rahaman:2021fcz} to make the density matrix separable when spin correlations are direct product of the polarizations, as done for the case of $tW^-$ before. 
The polarization and spin correlation parameters are computed from some asymmetries based on the joint angular distributions of the final state leptons at the rest frame of the two vector bosons after $VV$ system are Lorentz transformed to the center of mass frame.
The vector-vector, and vector-tensor relation to the asymmetries are given in the previous sections, while the tensor-tensor correlations are given by~\cite{Rahaman:2021fcz},
\begin{eqnarray}\label{eq:asym-TT}
	TT^{AB}_{(ij)(kl)} &=& \frac{2}{3}\left(\frac{3\pi}{4}\right)^2\frac{1}{(1-3\delta_A)}\frac{1}{(1-3\delta_B)}\frac{\sigma(c_i^ac_j^ac_k^bc_l^b>0)-\sigma(c_i^ac_j^ac_k^bc_l^b<0)}{\sigma(c_i^ac_j^ac_k^bc_l^b>0)+\sigma(c_i^ac_j^ac_k^bc_l^b<0)},\left(i\neq j, k\neq l\right) , \nonumber\\
	TT^{AB}_{(ij)zz} &=& \frac{8\pi^2}{3}\frac{1}{(1-3\delta_A)}\frac{1}{(1-3\delta_B)}\frac{\sigma(c_i^ac_j^ac_{zz}^b>0)-\sigma(c_i^ac_j^ac_{zz}^b<0)}{\sigma(c_i^ac_j^ac_{zz}^b>0)+\sigma(c_i^ac_j^ac_{zz}^b<0)},\left(i\neq j\right), \nonumber\\
	TT^{AB}_{(x^2-y^2)zz}&=&\frac{8\pi}{3}\frac{1}{(1-3\delta_A)}\frac{1}{(1-3\delta_B)}\frac{\sigma(c_{xxyy}^ac_{zz}^b>0)-\sigma(c_{xxyy}^ac_{zz}^b<0)}{\sigma(c_{xxyy}^ac_{zz}^b>0)+\sigma(c_{xxyy}^ac_{zz}^b<0)}, \nonumber\\
	TT^{AB}_{zzzz}&=&\frac{128}{27}\frac{1}{(1-3\delta_A)}\frac{1}{(1-3\delta_B)}\frac{\sigma(c_{zz}^ac_{zz}^b<0)-\sigma(c_{zz}^ac_{zz}^b<0)}{\sigma(c_{zz}^ac_{zz}^b<0)-\sigma(c_{zz}^ac_{zz}^b<0)}, \nonumber\\
	TT^{AB}_{(x^2-y^2)(x^2-y^2)}&=&\frac{3\pi^2}{2}\frac{1}{(1-3\delta_A)}\frac{1}{(1-3\delta_B)}\frac{\sigma(c_{xxyy}^ac_{xxyy}^b>0)-\sigma(c_{xxyy}^ac_{xxyy}^b<0)}{\sigma(c_{xxyy}^ac_{xxyy}^b>0)+\sigma(c_{xxyy}^ac_{xxyy}^b<0)}, \nonumber\\
	TT^{AB}_{(ij)(x^2-y^2)} 
	&=&\left(\frac{3\pi^2}{4}\right)\frac{1-3\delta_A}{1-3\delta_B}\frac{\sigma(c_i^ac_j^ac_{xxyy}^b>0)-\sigma(c_i^ac_j^ac_{xxyy}^b<0)}{\sigma(c_i^ac_j^ac_{xxyy}^b>0)+\sigma(c_i^ac_j^ac_{xxyy}^b<0)},\quad (i\neq j).
\end{eqnarray}
Here, $c_{xxyy}=c_x^2-c_y^2$ and $c_{zz}=\sin(3\theta)$, while $c_i$s are given Eq.~(\ref{eq:af-cxcycz}). The above tensor-tensor spin correlations along with their mirrors, i.e., $TT^{AB}_{zz(ij)}$,  $TT^{AB}_{zz(x^2-y^2)}$, and $TT^{AB}_{(x^2-y^2)(ij)}$ which can be obtained in similar manner, complete the parallelization of full density matrix of the $AB$ state.  At the SM leading order the polarizing power~($\alpha$) of lepton(anti-lepton) in the case of $W^{\mp}$ decay from Eqn.~(\ref{eqn:walpha}) is +1(-1) and the parameter $\delta$ is zero assuming the final fermions to be massless at high energy limits. 

We calculate all the asymmetries followed by all the polarization and spin correlations to fully reconstruct the production density matrices $\rho_{VV}$ in all three $VV$ production processes.   The measurement of the asymmetries is performed at the center of the mass frame of the $VV$ system after $V$ are Lorentz boosted to the rest frame. We choose the $z$-axis as the boost direction of visible particles. We do not consider the neutrinos to decide the $z$-axis for simplicity, though they can be reconstructed using various techniques like MAOS~\cite{Rahaman:2023pte} and ML algorithms~\cite{Raine:2023fko}. 
With the reconstructed density matrix from the polarization and spin correlation of the $VV$ state, we calculate the logarithmic negativity  ${\cal E}_N= \text{Log}_3(||\rho_{VV}^T||)$ define in Eq.~(\ref{eq:logdrho})  with $d=3$ as well as the lower bound of concurrence defined in Eqn.~(\ref{eq:em-low-diff-dim}). We estimate MC error to the entanglement measures using the similar procedure used in $t\bar{t}$ and $tW^-$ processes.  

\begin{table}[!b]
	\centering
	\caption{\label{tab:pponeone}
		The lower bound of concurrence and the logarithmic negativity are listed for the di-boson production processes~$(W^+W^-$, $ZZ$, and $ZW^+)$ at the $13$ TeV LHC along with 1$\sigma$ MC error. The average number of events used to estimate the measures are also listed in the second column.}
	\renewcommand{\arraystretch}{1.5}
	\begin{tabular*}{\textwidth}{@{\extracolsep{\fill}}lccccccccc@{}}\hline			
		\multirow{2}{*}{Bins} & \multicolumn{3}{c}{$W^+W^-$}&\multicolumn{3}{c}{$ZZ$}&\multicolumn{3}{c}{$ZW^+$}\\ \cline{2-4} \cline{5-7} \cline{8-10}
		&Events&$\mathcal{E}_N(\rho)$&$\mathcal{C}[\rho]_{LB}$&Events&$\mathcal{E}_N(\rho)$&$\mathcal{C}[\rho]_{LB}$&Events&$\mathcal{E}_N(\rho)$&$\mathcal{C}[\rho]_{LB}$\\\hline
		Unbin&$200000$&$0.232\pm0.013$&$0.170\pm0.010$&$200000$&$0.014\pm0.010$&$0.009\pm0.007$&$200000$&$0.006\pm0.007$&$0.004\pm0.005$\\
		M$_1$&$64083$&$0.262\pm0.022$&$0.205\pm0.020$&$31162$&$0.112\pm0.028$&$0.076\pm0.020$&$19404$&$0.094\pm0.033$&$0.063\pm0.024$\\
		M$_2$&$87098$&$0.261\pm0.018$&$0.202\pm0.016$&$113274$&$0.056\pm0.015$&$0.037\pm0.010$&$98143$&$0.031\pm0.015$&$0.020\pm0.010$\\
		M$_3$&$27024$&$0.285\pm0.030$&$0.220\pm 0.027$&$32013$&$0.100\pm0.031$&$0.068\pm0.022$&$42800$&$0.054\pm0.021$&$0.036\pm0.014$\\
		M$_4$&$10724$&$0.385\pm0.045$&$0.361\pm0.063$&$11935$&$0.247\pm0.044$&$0.181\pm0.037$&$18435$&$0.155\pm0.036$&$0.108\pm0.027$\\
		M$_5$&$11075$&$0.361\pm0.041$&$0.373\pm0.058$&$11620$&$0.239\pm0.043$&$0.174\pm0.036$&$21222$&$0.174\pm0.036$&$0.122\pm0.028$\\
		\hline
	\end{tabular*}
\end{table}
\begin{figure}[!h]
	\centering
	\includegraphics[width=0.4\textwidth]{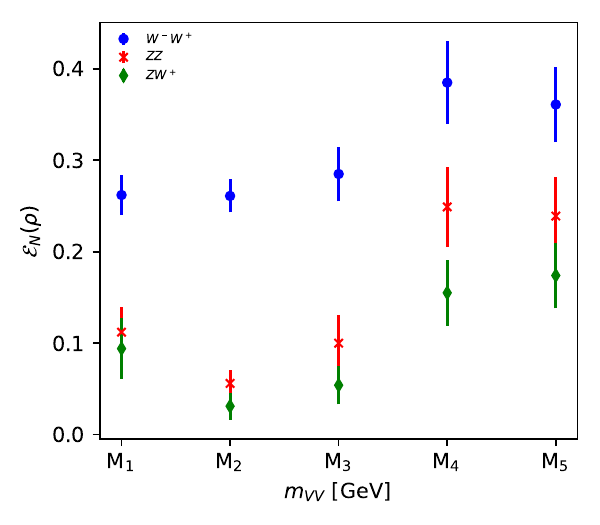}
	\includegraphics[width=0.4\textwidth]{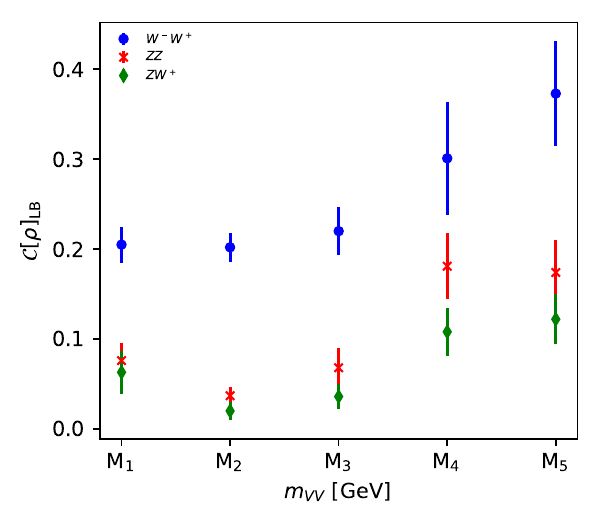}
	\caption{\label{fig:pponeone}Graphical visualization for the  measure of entanglement via logarithmic negativity ({\em left panel}) and lower bound of concurrence ({\em right panel}) for the di-boson production processes ($W^+W^-$, $ZZ$ and $W^+Z$) at the $13$ TeV LHC. The error bars correspond to 1$\sigma$ MC error.}
\end{figure}
For the LHC, the lower bound and  logarithmic negativity are calculated  in a few phase-space regions of di-boson invariant mass $m_{VV}$~(GeV) as,
\begin{equation}
	\begin{aligned}
		&\text{M}_1\equiv m_{VV}\le200,\quad\text{M}_2\equiv200<m_{VV}\le300,\quad\text{M}_3\equiv 300<m_{VV}\le400,\\
		&\text{M}_4\equiv 400<m_{VV}\le500,\quad \text{M}_5\equiv m_{VV}>500.
	\end{aligned}
\end{equation}
and they are listed in Table~\ref{tab:pponeone} along with the 1$\sigma$ MC standard deviation for all three di-boson production processes. The average number of events used to estimate the entanglement measures are also listed in the same table in the second column. 

In the case of $W^-W^+$ states, both measures of entanglement, i.e., $\mathcal{E}_N$ and $\mathcal{C}[\rho]_{\rm LB}$, imply an entangled $W^+W^-$ state for all chosen regions of $m_{W^+W^-}$, including the  full phase-space region (Unbin). The two measures do not differ by significant deviation. 
Moving on to the $ZZ$ state, both measures imply that the $ZZ$ state is entangled for regions of $m_{ZZ}$ and unbin region. However, the lower bound is not statistically significant for the unbin case, and it increases as invariant mass increases. 
For the $ZW^+$ state, both measures of entanglement are statistically zero implying a separable state in the unbin case. However, the measure of entanglement increases as one moves towards the higher invariant mass of $ZW^+$.  
The increase in the entanglement measure over the di-boson invariant mass is also demonstrated in the study by~\cite{Fabbrichesi:2023cev}, where they also consider the production angle of $V$.   For visual presentation, both measures of entanglement are shown in Fig.~\ref{fig:pponeone} as a function of invariant mass for all three di-boson processes considered. Notably, the $W^+W^-$ state is more entangled in comparison to $ZZ$ and $ZW^+$ in each bin of invariant mass, as prominently observed in the figure.

We also estimate the measure of entanglement for $e^+-e^-$ collider for the possible di-boson production processes $W^+W^-$, and $ZZ$ with un-polarized as well as polarized initial beams with $\sqrt{s}=1$ TeV. In this case, we divide the phase-space region with respect to the production angle of the gauge boson in a manner to maintain uniform statistics in all bins. It would prevent the over dependence of measurement on statistical fluctuations in different bins. The angular bins are chosen to be,
\begin{equation}
	\label{eqn:eeVV_window}
	\begin{aligned}
		{\rm C}_1 & \equiv |\cos\theta_V| \le 0.50, \quad {\rm C}_2 \equiv 0.50 < |\cos\theta_V| \le 0.70,
		{\rm C}_3 \equiv 0.70 < |\cos\theta_V| \le 0.80,\\ \quad {\rm C}_4 &\equiv 0.80 < |\cos\theta_V| \le 0.90,
		{\rm C}_5 \equiv 0.90 < |\cos\theta_V| \le 0.95, \quad {\rm C}_6 \equiv 0.95 < |\cos\theta_V| \le 1.0,
	\end{aligned}
\end{equation}
where $\theta_V$ is the production angle of gauge bosons with respect to the colliding beams.  We consider  $(\mp0.8,\pm0.3)$ as the longitudinal polarization for $(e^-,e^+)$ beams as benchmarks.
\begin{table}[!htb]
	\centering
	\caption{\label{tab:eeoneone}
		The lower bound of concurrence and the logarithmic negativity are listed for the $W^+W^-$,  and $ZZ$ at $e^+-e^-$ collider with $\sqrt{s}=1$~TeV  along with 1$\sigma$ MC error. The average number of events used to estimate the measures are also listed in the third column. The benchmark value of initial beam polarization used are $(\mp0.8,\pm0.3)$. }
	\renewcommand{\arraystretch}{1.5}
	\begin{tabular*}{\textwidth}{@{\extracolsep{\fill}}clcccccc@{}}\hline			
		\multirow{2}{*}{$(\eta_3,\xi_3)$} &\multirow{3}{*}{Bins}& \multicolumn{3}{c}{$W^+W^-$}&\multicolumn{3}{c}{$ZZ$} \\ \cline{3-5} \cline{6-8}
		&&Events&$\mathcal{E}_N(\rho)$&$\mathcal{C}[\rho]_{LB}$&Events&$\mathcal{E}_N(\rho)$&$\mathcal{C}[\rho]_{LB}$\\\hline
		&Unbin&$200000$&$0.150\pm0.015$&$0.133\pm0.034$&$200000$&$0.049\pm0.011$&$0.032\pm0.007$\\
		&C$_1$&$10621$&$0.852\pm0.028$&$0.978\pm0.052$&$16221$&$0.403\pm0.039$&$0.515\pm0.085$\\
		&C$_2$&$8602$&$0.636\pm0.041$&$0.720\pm0.067$&$11697$&$0.267\pm0.044$&$0.199\pm0.037$\\
		$(0.0,0.0)$&C$_3$&$7816$&$0.511\pm0.047$&$0.572\pm0.079$&$9589$&$0.263\pm0.042$&$0.194\pm0.036$\\
		&C$_4$&$15228$&$0.371\pm 0.041$&$0.383\pm0.065$&$17096$&$0.202\pm0.036$&$0.144\pm0.029$\\
		&C$_5$&$16623$&$0.313\pm0.037$&$0.302\pm0.065$&$17472$&$0.208\pm0.032$&$0.149\pm0.026$\\
		&C$_6$&$141114$&$0.123\pm0.016$&$0.129\pm0.043$&$127931$&$0.071\pm0.015$&$0.047\pm0.010$\\
		\hline
		&Unbin&$200000$&$0.155\pm0.014$&$0.133\pm0.031$&$200000$&$0.054\pm0.011$&$0.035\pm0.008$\\
		&C$_1$&$10372$&$0.862\pm0.028$&$1.010\pm0.051$&$16206$&$0.398\pm0.038$&$0.502\pm0.086$\\
		&C$_2$&$8537$&$0.611\pm0.044$&$0.704\pm0.073$&$11638$&$0.264\pm0.043$&$0.196\pm0.036$\\
		$(-0.8,+0.3)$&C$_3$&$7788$&$0.512\pm0.049$&$0.579\pm0.077$&$9542$&$0.274\pm0.045$&$0.204\pm0.039$\\
		&C$_4$&$15230$&$0.382\pm0.043$&$0.415\pm0.070$&$17117$&$0.199\pm0.035$&$0.142\pm0.028$\\
		&C$_5$&$16637$&$0.327\pm0.039$&$0.342\pm0.069$&$17563$&$0.207\pm0.033$&$0.148\pm0.027$\\
		&C$_6$&$141439$&$0.116\pm0.016$&$0.106\pm0.035$&$127940$&$0.069\pm0.014$&$0.046\pm0.010$\\
		\hline
		&Unbin&$200000$&$0.136\pm0.014$&$0.102\pm0.022$&$200000$&$0.059\pm0.012$&$0.038\pm0.008$\\
		&C$_1$&$14169$&$0.686\pm0.031$&$0.718\pm0.055$&$16165$&$0.398\pm0.037$&$0.528\pm0.086$\\
		&C$_2$&$9480$&$0.576\pm0.044$&$0.669\pm0.071$&$11659$&$0.292\pm0.046$&$0.220\pm0.040$\\
		$(+0.8,-0.3)$&C$_3$&$8023$&$0.494\pm0.051$&$0.561\pm0.081$&$9570$&$0.277\pm0.049$&$0.206\pm0.042$\\
		&C$_4$&$14998$&$0.350\pm 0.041$&$0.378\pm0.070$&$17169$&$0.206\pm0.035$&$0.147\pm0.028$\\
		&C$_5$&$16264$&$0.315\pm0.037$&$0.287\pm0.061$&$17538$&$0.198\pm0.036$&$0.141\pm0.028$\\
		&C$_6$&$137071$&$0.117\pm0.016$&$0.094\pm0.028$&$127903$&$0.073\pm0.014$&$0.049\pm0.010$\\
		\hline
	\end{tabular*}
\end{table}
\begin{figure}[!htb]
	\centering
	\includegraphics[width=0.4\textwidth]{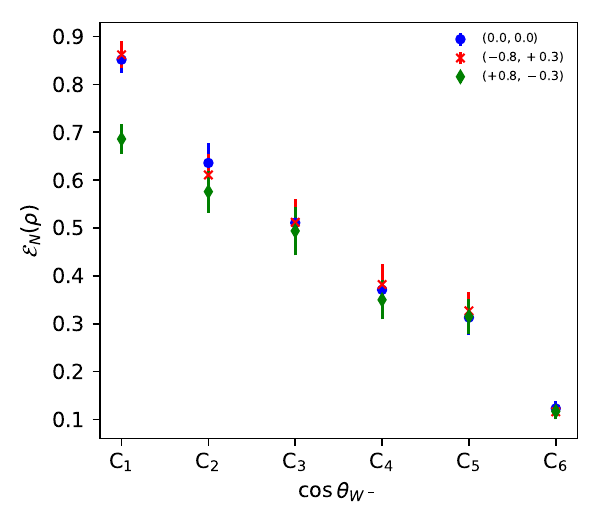}
	\includegraphics[width=0.4\textwidth]{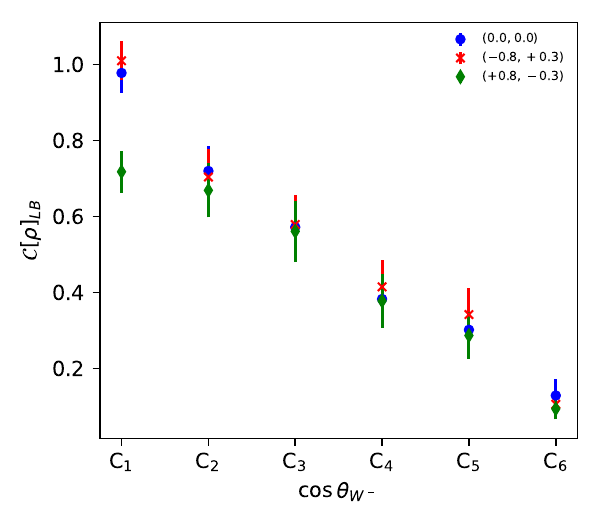}
	\includegraphics[width=0.4\textwidth]{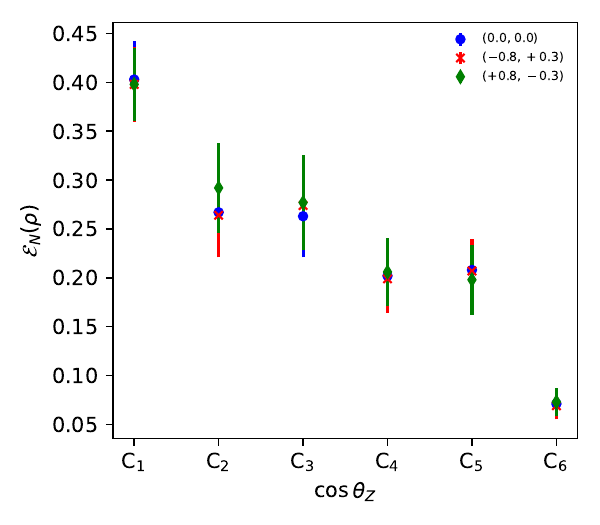}
	\includegraphics[width=0.4\textwidth]{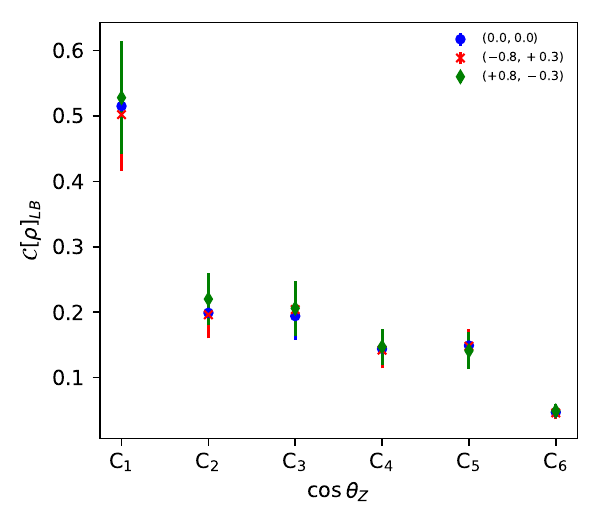}
	\caption{\label{fig:eeoneone}Graphical representation of measure of entanglement via logarithmic negativity and lower bound for $W^-W^+$ process ({\em top row}), and $ZZ$ process ({\em bottom-row}) for $e^-e^+$ collider with $\sqrt{s}=1$~TeV. The analysis is done for unpolarized beams along with two sets of longitudinal beam polarizations, $(\eta_3,\xi_3)=(\mp 0.8, \pm 0.3)$. The error bars correspond to 1$\sigma$ MC deviations.}
\end{figure} 
We list both the entanglement measures i.e, logarithmic negativity~$(\mathcal{E}_N$ and lower bound~$(\mathcal{C}[\rho]_{LB})$ on the entanglement of formation  in Table~\ref{tab:eeoneone} with $1\sigma$ deviation in measurement for unpolarized and polarized initial beams. The average number of events in each region of the production angle are also given in the same table for both di-boson processes. For both di-boson processes, the two bosons are inferred to be entangled from both measures of entanglement for all chosen regions of production angle for all three sets of beam polarization.
We note that both measure undergoes linear reduction as a function of the cosine of the production angle of $W$ boson in the $W^-W^+$ process.  The measures reduce as one moves from $(|\cos\theta_W| = 0.0)$ to $(\cos\theta_W = +1.0)$ region. A similar behavior is also observed for the $ZZ$ process in the case of $e^+e^-$ collider. For a visual representation,  the two measures of entanglement are depicted in Fig.~(\ref{fig:eeoneone}) for both  $W^+W^-$ and $ZZ$ processes. For both the di-boson process, both measures of entanglement do not differ much in un-polarized and $(-0.8,+0.3)$ polarization of beams, while they are comparatively smaller for the reverse polarizations of beams, $(+0.8,-0.3)$.

\section{Tripartite entanglement  in $t\bar{t}Z$ production}
\label{sec:halfhalfone}
In this section, we explore multi-partite quantum entanglement in $\ttz$ production process at the LHC as an example of $2\otimes2\otimes3$ spin quantum state beyond the simplest tripartite case which is $2\otimes2\otimes2$. Similar to the previous bi-partite cases, we estimate various entanglement measures from the reconstructed density matrix of the $\ttz$ state from polarizations and all possible spin correlations measured by analyzing the angular distributions of the top quarks and the $Z$ bosons in MC simulations. The full spin correlated  density matrix for the $t\bar{t}Z$ production process is represented by~\cite{Rahaman:2022dwp},
\begin{eqnarray}\label{eq:spin-density-half-half-one}
	\rho_{\ttz}
	&=& \frac{1}{12}
	\Big[ \mathbb{I}_{12\times 12}  
	+ \vec{p}^t\cdot\vec{\sigma}\otimes \mathbb{I}_{6\times 6}
	+\mathbb{I}_{2\times 2}\otimes\vec{p}^{\tb}\cdot\vec{\sigma}\otimes \mathbb{I}_{3\times 3} \nonumber\\
	&+& \frac{3}{2}\mathbb{I}_{4\times 4} \otimes \vec{p}^Z\cdot\vec{S}
	+\sqrt{\frac{3}{2}} \mathbb{I}_{4\times 4} \otimes  T_{ij}^Z\big(S_iS_j+S_jS_i\big)    
	+ pp_{ij}^{t\tb}\sigma_i\otimes \sigma_j \otimes \mathbb{I}_{3\times 3}  \nonumber\\
	&+& \frac{3}{2}pp_{ij}^{tZ}\sigma_i\otimes\mathbb{I}_{2\times 2}\otimes S_j + \sqrt{\frac{3}{2}} pT_{ijk}^{tZ}\sigma_i\otimes\mathbb{I}_{2\times 2}\otimes (S_jS_k+S_kS_j) \nonumber\\
	&+& \frac{3}{2}pp_{ij}^{\tb Z}\mathbb{I}_{2\times 2}\otimes\sigma_i\otimes S_j 
	+\sqrt{\frac{3}{2}}  pT_{ijk}^{\tb Z}\mathbb{I}_{2\times 2}\otimes \sigma_i\otimes(S_jS_k+S_kS_j) \nonumber\\
	&+& \frac{3}{2}ppp_{ijk}^{\ttz}\sigma_i\otimes\sigma_j\otimes S_k +\sqrt{\frac{3}{2}} ppT_{ijkl}^{\ttz}\sigma_i\otimes\sigma_j\otimes (S_kS_l+S_lS_k)  \Big].
	%
\end{eqnarray} 
Here,  $\vec{p}^{t/\tb/Z}$ are the vector polarizations of $t/\tb/Z$; $T_{ij}^{Z}$ are the  tensor polarizations of $Z$; $pp^{AB}$ are the vector-vector spin correlations of $A$-$B$ ($A/B=t/\tb/Z$) pair;  $pT_{ijk}^{tZ}$ ($pT_{ijk}^{\tb Z}$) are the  vector-tensor spin correlations of $t$-$Z$ ($\tb$-$Z$) pair; $ppp_{ijk}^{\ttz}$ and $ppT_{ijkl}^{\ttz}$ are the vector-vector-vector  and vector-vector-tensor three body spin correlations, respectively of $t$-$\tb$-$Z$ system. The $pT_{i(jk)}^{(t/\bar{t})Z}$ and $ppT_{ij(kl)}^{\ttz}$ are symmetric in the last two indices similar to the $T^Z$~\cite{Rahaman:2021fcz}. The correlations $T$, $pT$, and $ppT$ have zero traces in the indices of $T$, i.e.,
\begin{equation}
	\sum_iT_{ii}=0,~\sum_jpT_{ijj}=0,~\sum_k ppT_{ijkk}=0.
\end{equation}
We introduced factors such as $3/2$ and $\sqrt{3/2}$ in Eq.~(\ref{eq:spin-density-half-half-one})against Ref.~\cite{Rahaman:2022dwp} for the same purpose we did for $tW^-$ and $VV$ process in previous sections. 
The polarization and spin correlation parameters can be measured from the angular distributions of the decay products of the $t$, $\tb$, and $Z$ at the reset frame of their mother particles before Lorentz boosting to the $\ttz$ c.m. frame. In the fully leptonic final state, 
the three-body spin correlations can be obtained from the following angular asymmetries~\cite{Rahaman:2022dwp},
\begin{eqnarray}\label{eq:asym-App-half-one}
	{\cal A}\left[ppp_{ijk}^{\ttz}\right] 
	&=&\dfrac{\sigma\l(c_i^{l_t} c_j^{l_{\tb}} c_k^{l_Z}>0\r)-\sigma\l(c_i^{l_t} c_j^{l_{\tb}} c_k^{l_Z}<0\r)}
	{\sigma\l(c_i^{l_t} c_j^{l_{\tb}} c_k^{l_Z}>0\r)+\sigma\l(c_i^{l_t} c_j^{l_{\tb}} c_k^{l_Z}<0\r)},\nonumber\\
	&=&\frac{3}{16}\alpha_t\alpha_{\tb} \alpha_Z ~ppp_{ijk}^{\ttz},\nonumber\\
	{\cal A}\left[ppT_{ij(kl)}^{AB}\right] 
	&=&\dfrac{\sigma\l(c_i^{l_t} c_j^{l_{\tb}} c_k^{l_Z} c_l^{l_Z}>0\r)-\sigma\l(c_i^{l_t} c_j^{l_{\tb}c_l^{l_Z}} c_k^{l_Z}<0\r)}
	{\sigma\l(c_i^{l_t} c_j^{l_{\tb}} c_k^{l_Z} c_l^{l_Z}>0\r)+\sigma\l(c_i^{l_t} c_j^{l_{\tb}c_l^{l_Z}} c_k^{l_Z}<0\r)},~(k\ne l), \nonumber\\
	&=& \sqrt{\frac{3}{2}}\frac{1}{6\pi}\alpha_t\alpha_{\tb}(1-3\delta_B) ppT_{ij(kl)}^{\ttz},\nonumber\\ 
	{\cal A}\left[ppT_{ij(x^2-y^2)}^{\ttz}\right] 
	&=&\dfrac{\sigma\l(c_i^{l_t} c_j^{l_{\tb}} \l((c_{x}^{l_Z})^2-(c_{y}^{l_Z})^2\r) >0\r)-\sigma\l(c_i^{l_t} c_j^{l_{\tb}} \l((c_{x}^{l_Z})^2-(c_{y}^{l_Z})^2\r) <0\r)}{\sigma\l(c_i^{l_t} c_j^{l_{\tb}} \l((c_{x}^{l_Z})^2-(c_{y}^{l_Z})^2\r) >0\r)+\sigma\l(c_i^{l_t} c_j^{l_{\tb}} \l((c_{x}^{l_Z})^2-(c_{y}^{l_Z})^2\r) <0\r)},\nonumber\\
	&=& \sqrt{\frac{3}{2}}\frac{1}{6\pi}\alpha_t\alpha_{\tb}(1-3\delta_B) ppT_{ij(x^2-y^2)}^{\ttz},\nonumber\\ 
	{\cal A}\left[ppT_{ijzz}^{\ttz}\right] 
	&=& \dfrac{\sigma\l(c_i^{l_t} c_j^{l_{\tb}} \sin\l(3\theta_{l_Z}\r)>0\r)-\sigma\l(c_i^{l_t} c_j^{l_{\tb}} \sin\l(3\theta_{l_Z}\r)<0\r)}{\sigma\l(c_i^{l_t} c_j^{l_{\tb}} \sin\l(3\theta_{l_Z}\r)>0\r)+\sigma\l(c_i^{l_t} c_j^{l_{\tb}} \sin\l(3\theta_{l_Z}\r)<0\r)}  ,\nonumber\\
	&=& \sqrt{\frac{3}{2}}\frac{3}{32}\alpha_t\alpha_{\tb}(1-3\delta_B) ppT_{ijzz}^{\ttz}.
\end{eqnarray}
Here, $l_A$ denotes lepton decayed from the particle $A$. The asymmetries for the polarization and various two-body spin correlations are discussed in $t\bar{t}$ and $tW^-$ process in previous sections.


\begin{figure}[h!]
	\centering
	\includegraphics[width=0.6\textwidth]{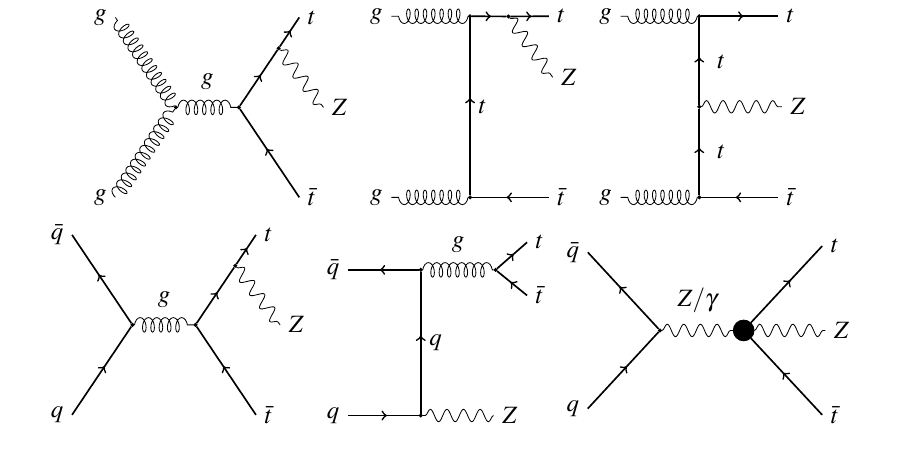}
	\caption{\label{fig:feynman-ttz} Representative Feynman diagrams for the production $\ttz$ process in the SM  and  with the $t\bar{t}Z(Z/\gamma)$ quartic vertex (represented by the shaded blob in the {\em right-bottom} diagram) generated by the dim-$8$ effective operator given in Eq.~(\ref{eq:dim8-operator}).}
\end{figure}
\begin{figure}[htb!]
	\centering		
	\includegraphics[width=0.4\textwidth]{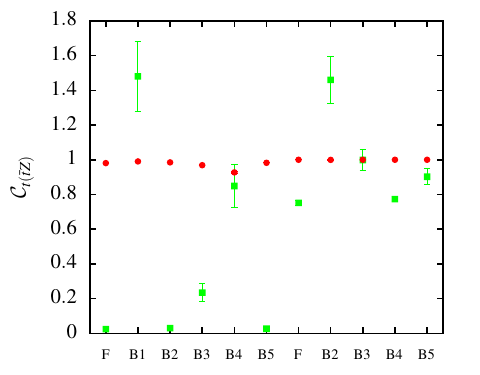}	
	\includegraphics[width=0.4\textwidth]{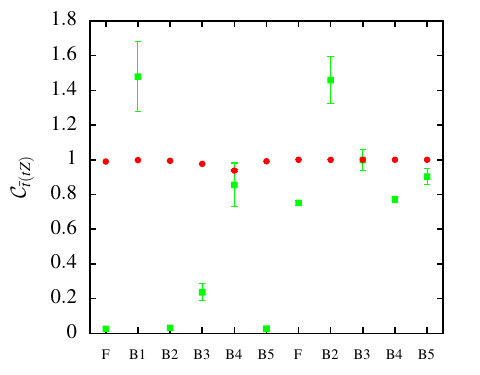}	
	\includegraphics[width=0.4\textwidth]{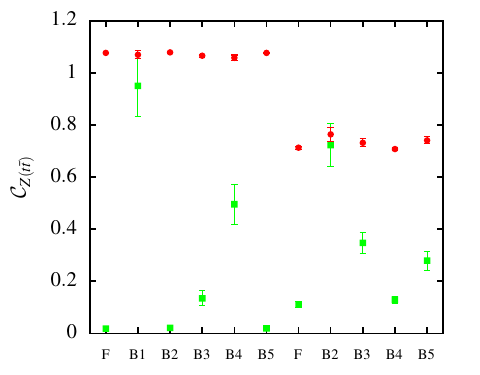}	
	\includegraphics[width=0.4\textwidth]{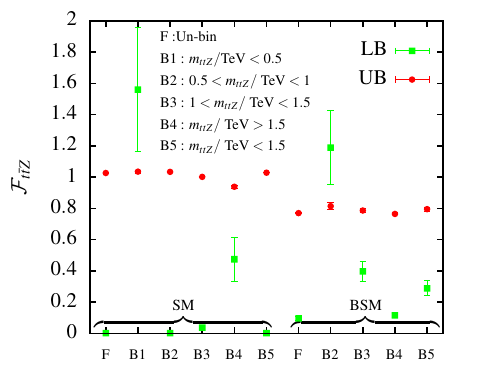}		
	\caption{\label{fig:QEM-ttz-set1} Lower and upper bounds on the  one-to-other bipartite and tripartite entanglement are presented  for $\ttz$ process at the $13$ TeV LHC  in SM and BSM in different $m_{\ttz}$ region. The error bars correspond to $1\sigma$ Monte-Carlo. }
\end{figure}
In addition to the SM, we explore various entanglement measures in the $\ttz$  process at the LHC in a BSM scenario.  In the BSM scenario, we work with dimension-$8$ operators of the form~\cite{Murphy:2020rsh},
\begin{eqnarray}\label{eq:dim8-operator}
	{\cal O}_{quW^2H}=\l(\bar{q}_p u_r\r)\wtil{H}W_{\mu\nu}^aW^{a\mu\nu},
\end{eqnarray}
where $p$, $r$ are the quark flavor indices and $\wtil{H}_j=\epsilon_{ij}H^{\dagger k}$. This operator provides a quartic vertex of $(\gamma/Z)Zt\bar{t}$ after the Higgs boson gets vacuum expectation value. This operator is added to the SM Lagrangian as,
\begin{eqnarray}
	{\cal L}_{\rm SM} + \frac{c_{quW^2H}}{\Lambda^4} {\cal O}_{quW^2H},
\end{eqnarray}
where   $c_{quW^2H}$ is the Wilson coefficient and $\Lambda$ is a cutoff scale. 
We simulate the $\ttz$ process at the $13$ TeV LHC in both SM and the BSM case in the fully leptonic, i.e., four leptons, two $b$-jet, and missing transverse energy final state. For BSM case, we choose $\Lambda=1$ TeV and  $c_{quW^2H}=1$.  The representative Feynman diagrams for the $\ttz$ production are given in Fig.~\ref{fig:feynman-ttz} where the {\em right-bottom} diagram with shaded blob correspond to the $t\bar{t}Z(Z/\gamma)$ quartic vertex generated by the dim-$8$ effective operator given in Eq.~(\ref{eq:dim8-operator}). The rest of the diagrams correspond to the SM process.

We analyze parton-level events using truth-level neutrino momenta to derive the polarizations and spin correlations of the $\ttz$ process in the center-of-mass frame. The missing neutrinos, however, can be reconstructed using the on-shell mass approximation of the top quarks and the $Z$ boson~\cite{Rahaman:2022dwp}.
We generate 1{\tt M} events both in SM and  BSM at the decay level and estimate the polarization and spin correlation parameters using angular distributions of the leptons.  
We estimate the lower and upper bound of entanglement for all possible quantum states, including the bipartite $t\bar{t}$, $tZ$, and $\bar{t}Z$ states, as well as the one-to-other bipartite states, i.e., $t|\bar{t}Z$, $\bar{t}|tZ$, and $Z|t\bar{t}$.  The one-to-other bipartite entanglements are subsequently utilized to estimate the tripartite entanglement through the area of concurrence triangle, as defined in Eq.~(\ref{eq:GME-F}).  The trace squares of different states, such as single, bipartite, and tripartite, in the $\ttz$ process, necessary for the lower bounds, are provided in Appendix\ref{sec:tr-rho-sq}. Some of these trace squares are also applicable to the $t\bar{t}$ and $tW^-$ processes.

Entanglement measures are computed for various reconstructed $m_{\ttz}$ regions, including the unbin case, in both SM and BSM scenarios. These measures, along with 1$\sigma$ Monte Carlo errors estimated by selecting $30\%$ of events in 1000 iterations, are listed in Table~\ref{tab:em-ttz}. Additionally, the average number of events used for the entanglement measures is provided in each bin and unbin region in the second column of the table. Finally, the entanglement measures, including lower and upper bounds, are graphically depicted in Fig.~\ref{fig:QEM-ttz-set1}.

We observe no entanglement in the $t\bar{t}$ state across all regions of invariant mass in the SM. Similarly, the other bipartite states, namely $tZ$ and $\bar{t}Z$, also exhibit no entanglement across all bins, except for $\sqrt{\hat{s}} < 0.5$ and $\sqrt{\hat{s}} < 1.5$ TeV, with low significance in the latter bin. The one-to-other bipartite entanglements are nearly zero in the unbin case, leading to near-zero tripartite entanglement. While we obtain a lower bound greater than 1 (the lower bound exceeds the upper bound, as depicted in Fig.~\ref{fig:QEM-ttz-set1}) for the values of one-to-other bipartite entanglement in the region $\sqrt{\hat{s}} < 0.5$ and hence the tripartite entanglement, this is attributed to large errors on asymmetries due to the low number of events in this bin. Tripartite entanglement is observed (at $47\%$) at higher invariant masses with $\sqrt{\hat{s}} > 1.5$ TeV, while it remains negligible in other bins.

In the context of BSM, the $t\bar{t}$ state shows entanglement (ranging from $66\%$ to $71\%$) across all regions of the invariant mass, contrasting with the SM case. However, $tZ$ and $\bar{t}Z$ states demonstrate near-zero entanglement, similar to the SM case. Tripartite entanglement is present in all bins of the invariant mass, with the least value observed in the unbin case. In the BSM scenario, the entanglement measure is unavailable in the $0.5 < \sqrt{\hat{s}} < 1$ TeV bin, as practically no events are present due to the momentum dependence of the dimension-$8$ operator, causing events to shift towards higher invariant masses. The entanglement measures are more reliable (with low Monte Carlo error) in the higher bins in the BSM case compared to the SM case, owing to larger event samples. The operator induces some level of entanglement, particularly in the $t\bar{t}$ and tripartite states.

It is worth noting that the total number of events, and hence the events in the bins, are chosen to be considerably large compared to the expected number of events at the LHC~\cite{CMS:2019too}. Therefore, the detectability of tripartite entanglement in the $\ttz$ process is not guaranteed at the LHC. This study aims to provide a conceptual understanding of genuine multipartite entanglement phenomena in this process. However, one can expect to observe this tripartite entanglement in the SM at the Future Circular Collider (FCC)~\cite{FCC:2018vvp}.
\begin{sidewaystable}
	\caption{\label{tab:em-ttz} The lower bound for the different measures of entanglement in $\ttz$ process in SM and BSM (with dimension-$8$ effective operator) estimated using Eq.~(\ref{eq:em-low-diff-dim}) for bipartite entanglement. For tripartite entanglement in the last two columns represented by superscript $(1) $ and $(2)$, we use the method outlined in Eqs.~(\ref{eq:GME-F}) and (\ref{eq:GME2}), respectively.}
	\renewcommand{\arraystretch}{1.5}
	\begin{tabular*}{\textwidth}{@{\extracolsep{\fill}}clcccccccc@{}}\cline{2-10}		
		&Bin&Events&$t\tb $&$ tZ $&$ \tb Z$&$ t|\tb Z $&$ \tb|tZ $&$ Z|t\tb $&$ \ttz $\\\cline{2-10}
		&Unbin&300000&$0.0$&$0.0$&$0.0$&$0.024\pm0.011$&$0.026\pm0.012$&$0.018\pm0.008$&$0.001\pm0.001$\\\cline{2-10}
		&$\sqrt{\hat{s}}<0.5$ TeV&6002 &$0.0	$&$0.129\pm0.070$&$0.105\pm0.061$&$1.480\pm0.203$&$1.479\pm0.202$&$0.950\pm0.116$&$1.560\pm0.397$\\\cline{2-10}
		&$0.5<\sqrt{\hat{s}}<1$ TeV&220710&$0.0$&$0.0$&$0.0$&$0.030\pm0.016$&$0.031\pm0.016$&$0.021\pm0.009$&$0.001\pm0.001$\\\cline{2-10}
		&$1<\sqrt{\hat{s}}<1.5$ TeV&59364&$0.0$&$0.000\pm0.003$&$0.000\pm0.001$&$0.234\pm0.050$&$0.237\pm0.050$&$0.135\pm0.028$&$0.036\pm0.014$\\\cline{2-10}
		&$\sqrt{\hat{s}}>1.5$ TeV&13924&$0.0$&$0.053\pm0.037$&$0.059\pm0.039$&$0.849\pm0.125$&$0.855\pm0.126$&$0.496\pm0.076$&$0.474\pm0.139$\\\cline{2-10}
		\multirow{-6}{*}[4.5em]{\begin{sideways}$\overbrace{\hspace{4.4cm}}^{\text{SM}}$\end{sideways}} 
		&$\sqrt{\hat{s}}<1.5$ TeV&286076&$0.0$&$0.0$&$0.0$&$0.027\pm0.013$&$0.027\pm0.014$&$0.019\pm0.008$&$0.001\pm0.001$\\\cline{2-10}
		&Un-bin&300000&$0.708\pm0.005$&$0.000\pm0.001$&$0.002\pm0.003$&$0.752\pm0.013$&$0.751\pm0.013$&$0.111\pm0.013$&$0.096\pm0.012$\\\cline{2-10}
		&$0.5<\sqrt{\hat{s}}<1$ TeV&9923&$0.668\pm0.027$&$0.084\pm0.049$&$0.075\pm0.045$&$1.460\pm0.135$&$1.459\pm0.134$&$0.723\pm0.084$&$1.188\pm0.237$\\\cline{2-10}
		&$1<\sqrt{\hat{s}}<1.5$ TeV&36341&$0.695\pm0.015$&$0.017\pm0.016$&$0.012\pm0.015$&$0.999\pm0.061$&$0.998\pm0.061$&$0.348\pm0.041$&$0.396\pm0.065$\\\cline{2-10}
		&$\sqrt{\hat{s}}>1.5$ TeV&253724&$0.712\pm0.005$&$0.005\pm0.005$&$0.009\pm0.006$&$0.774\pm0.017$&$0.773\pm0.017$&$0.129\pm0.014$&$0.115\pm0.014$\\\cline{2-10}
		\multirow{-5}{*}[3.5em]{\begin{sideways}$\overbrace{\hspace{3.5cm}}^{\text{dim-$8$ BSM}}$\end{sideways}}  
		&$\sqrt{\hat{s}}<1.5$ TeV&46276&$0.688\pm0.013$&$0.004\pm0.008$&$0.003\pm0.007$&$0.902\pm0.047$&$0.902\pm0.047$&$0.279\pm0.036$&$0.288\pm0.048$\\\cline{2-10}
	\end{tabular*}	
\end{sidewaystable}

	\section{Summary}\label{sec:conclude}	
High-energy particle colliders such as the LHC and future colliders provide a natural platform for exploring quantum entanglement at relativistic limits. The recent observation of entanglement by ATLAS in $t\bar{t}$ production at the LHC has sparked significant interest in investigating entanglement phenomena across various processes at particle colliders. While entanglement phenomena are extensively studied in bipartite states such as $t\bar{t}$ and $WW/ZZ$ production processes in both the SM and beyond, tripartite entanglement remains relatively underexplored.

In this article, we delve into the phenomena of tripartite entanglement in $\ttz$ production at the LHC, while also considering bipartite entanglement in $t\bar{t}$, $tW^-$, and $WW/WZ/ZZ$ production processes. We first outline the methodology for testing and measuring entanglement in both bipartite and tripartite cases, employing a density matrix approach. We reconstruct the density matrix, which is parameterized by polarization and spin correlations, leveraging asymmetries derived from the angular distributions of the final-state leptons in the rest frame of the top quarks and the vector bosons.

In $t\bar{t}$ production, we compute four different measures of entanglement: the absolute sum of negative eigenvalues of the partially transposed matrix (${\cal C}^{\text{PPT}}$), concurrence ($\mathcal{C}(\rho)$), the correlation matrix (${\cal C}^\pm$), and $D = -3.0 \cdot \langle \cos\theta_{\ell\ell} \rangle$, and compare them. While some measures agree with each other, others do not. The $D$ marker indicates entangled $t\bar{t}$ states at lower invariant masses ($m_{t\bar{t}}$), while concurrence suggests the same for higher invariant masses at the LHC.

Moving on to $tW^-$ production, both PPT entanglement and the lower bound of concurrence indicate entanglement for higher invariant masses. In the di-boson production processes, we estimate the logarithmic negativity and lower bound of concurrence to explore quantum entanglement. Both measures suggest the presence of entanglement in all di-boson production processes ($WW$, $ZZ$, $WZ$) at the LHC and future $e^+e^-$ colliders across various regions of phase space, such as di-boson invariant mass for the LHC and production angle for $e^+e^-$ colliders.

Finally, we present our main result on tripartite entanglement in the $\ttz$ production process at the LHC. Tripartite entanglement is present at high $\ttz$ invariant masses ($\sqrt{\hat{s}}>1.5$ TeV) in both the SM and in BSM scenarios with a dimension-$8$ effective operator. We utilize parton-level simulated events and truth neutrino information for various measures of entanglement. These studies are intended to provide methodologies for probing both bipartite and tripartite entanglement through polarization and spin correlation of top quarks and heavy gauge bosons.
	
	\section*{Acknowledgment}
	Authors thanks Ritesh Singh for the initial encouragement to work on this article and for many fruitful discussions which had helped to shape the article in its current form. R.R is supported by FAPESP fellowship with grand 2023/04036-1. A. S. thanks the
	University Grant Commission, Government of India, for
	financial support through the UGC-NET Fellowship.
	\begin{appendices}
	\section{Different ${\rm Tr}(\rho^2)$ in $\ttz$ production}\label{sec:tr-rho-sq}
		Different quantities for the bipartite and tripartite entanglement are given bellow.
		\begin{equation}
			{\rm Tr}\l(\rho_{t/\tb}^2\r)=\frac{1}{2}\l[1+\sum_i \l(p_i^{t/\tb}\r)^2  \r]
		\end{equation}
		\begin{equation}
			{\rm Tr}\l(\rho_{t\tb}^2\r)=\frac{1}{4}\l[1+\sum_i \l(p_i^{t}\r)^2+\sum_i \l(p_i^{\tb}\r)^2
			+\sum_{i,j} \l(pp_{ij}^{t\tb}\r)^2  \r]
		\end{equation}
		
		\begin{equation}
			{\rm Tr}\l(\rho_{Z}^2\r)=\frac{1}{3}\l[1+\frac{3}{2}\sum_i \l(p_i^{Z}\r)^2 + \l(T^Z_{x^2-y^2}\r)^2+3\l(T_{33}^Z\r)^2 + \sum_{i,j}\l(T^Z_{ij}+T^Z_{ji}\r)^2(i\ne j)  \r]
		\end{equation}
		
		\begin{eqnarray}
			{\rm Tr}\l(\rho_{tZ}^2\r)=\frac{1}{6}\bigg[1 &+& \l(2 {\rm Tr}\l(\rho_{t}^2\r)-1 \r)	 + \l(3 {\rm Tr}\l(\rho_{Z}^2\r)-1 \r)	
			+\frac{3}{2}\sum_{i,j} \l(pp_{ij}^{tZ}\r)^2 \nonumber \\
			&+&\sum_i \l(pT^{tZ}_{i(x^2-y^2)}\r)^2+3\sum_i\l(pT_{i33}^{tZ}\r)^2 + \sum_{i,j,k}\l(pT^{tZ}_{ijk}+pT^{tZ}_{ikj}\r)^2(j\ne k)
			\bigg]
		\end{eqnarray}
		\begin{eqnarray}
			{\rm Tr}\l(\rho_{\ttz}^2\r)=\frac{1}{12}\bigg[1  &+&	\l(6{\rm Tr}\l(\rho_{tZ}^2\r)-1\r)
			+	\l(6{\rm Tr}\l(\rho_{\tb Z}^2\r)-1\r) +	\l(4{\rm Tr}\l(\rho_{t\tb}^2\r)-1\r) \nonumber\\
			&-& \l(2{\rm Tr}\l(\rho_{t}^2\r)-1\r)- \l(2{\rm Tr}\l(\rho_{\tb}^2\r)-1\r) - \l(3{\rm Tr}\l(\rho_{Z}^2\r)-1\r)\nonumber\\
			&+&\frac{3}{2}\sum_{i,j,k} \l(ppp_{ijk}^{\ttz}\r)^2    + \sum_{i,j}\l(ppT^{\ttz}_{ij(x^2-y^2)}\r)^2+3\sum_{i,j}\l(ppT_{ij33}^{\ttz}\r)^2 \nonumber\\
			&+& \sum_{i,j,k,l}\l(ppT^{\ttz}_{ijkl}+ppT^{\ttz}_{ijlk}\r)^2(k\ne l)
			\bigg]
		\end{eqnarray}
		
		Here we defined,
		\begin{eqnarray}
			T_{x^2-y^2}&=&T_{11}-T_{22}\nonumber\\
			pT_{i(x^2-y^2)}&=&pT_{i11}-pT_{i22}\nonumber\\
			ppT_{ij(x^2-y^2)}&=&ppT_{ij11}-ppT_{ij22}.	
		\end{eqnarray}
		
		The spin density matrices for the top quark and $Z$ are given by~\cite{Bourrely:1980mr,Boudjema:2009fz}
		\begin{eqnarray}
			\rho_t&=&\frac{1}{2}\l[\mathbb{I}_{2\times 2}+\vec{p}^t\cdot\vec{\sigma}\r],	\\
			\rho_Z&=&\frac{1}{3}\l[\mathbb{I}_{3\times 3}+\frac{3}{2}\vec{p}^Z\cdot\vec{S}  +\sqrt{\frac{3}{2}}   T_{ij}^Z\big(S_iS_j+S_jS_i\big)     \r].	
		\end{eqnarray}

	\end{appendices}
	
	
	\bibliography{sn-bibliography}

\end{document}